\documentclass[12pt]{article}
\usepackage{graphicx}
\usepackage{subfigure}
\usepackage{geometry}
\usepackage{latexsym}
\usepackage{amssymb, amstext, amsgen}
\usepackage{amsthm}

\theoremstyle{definition}

\makeatletter
\@addtoreset{equation}{section} 

\begin{document}
\baselineskip=18.6pt plus 0.2pt minus 0.1pt

\date{}
\begin{titlepage}
\title{
\hfill\parbox{4cm}
{\normalsize UFR-HEP/00/20}\\
\vspace{1cm}
{ \bf   Toric Geometry,  Enhanced non Simply laced Gauge Symmetries in  Superstrings and F-theory Compactifications}
}
\author{
A. Belhaj\thanks{{\tt ufrhep@fsr.ac.ma}} and E.H Saidi\thanks{{\tt  H-saidi@fsr.ac.ma}}
{}
\\[7pt]
{\it  High Energy Physics Laboratory and UFR, Faculty of sciences, Rabat, Morocco}
}
\maketitle
\thispagestyle{empty}
\def\[{\bigl[} \def\]{\bigr]} \def\({\bigl(} \def\){\bigr)} \def\p{\partial}
\def\o{\over} \def\ta{\tau} \def\cm{\cal M} \def\R{\bf R} \def\b{\beta} \def%
\a{\alpha} \def\be{\begin{equation}}
\def\ee{\end{equation}}
\def\bea{\begin{eqnarray}}
\def\eea{\end{eqnarray}}
\def\nn{\nonumber}

\begin{abstract}
 We study the geometric engineering of supersymmetric quantum field theories (QFT), with non simply laced gauge groups, obtained from superstring and F-theory compactifications on local Calabi-Yau manifolds. First we review the main lines of the toric method for ALE spaces with ADE singularities which we extend to non simply laced ordinary and affine singularities. Then, we develop two classes of solutions depending on the two possible realisations of the outer-automorphism group of the toric graph $\Delta$(ADE). In F-theory on elliptic Calabi-Yau manifolds, we give explicit results for the affine non simply laced toric data and the corresponding BCFG mirror geometries. The latters extend known results obtained in litterature for the affine ADE cases. We also study the geometric engineering of $ N=1$ supersymmetric gauge theory in eight dimensions. In type II superstring compactifications on local Calabi-Yau threefolds, we complete the analysis for ordinary ADE singularities by giving the explict derivation of the lacking non simply ones. Finally we develop the basis of polyvalent toric geometry. The latter extends  bivalent and trivalent geometries, considered in the geometric engineering method, and use it to derive a new solution for the affine $\hat  D_4$ singularity. Other features are also discussed.
\end{abstract}
\end{titlepage}
\newpage
\tableofcontents

\newpage
\newpage
\def\be{\begin{equation}}
\def\ee{\end{equation}}
\def\bea{\begin{eqnarray}}
\def\eea{\end{eqnarray}}
\def\nn{\nonumber}
\def\t{\times}
\def\[{\bigl[}
\def\]{\bigr]}
\def\({\bigl(}
\def\){\bigr)}
\def\p{\partial}
\def\o{\over}
\def\ta{\tau}
\def\cm{\cal M}
\def\R{\bf R}
\def\b{\beta}
\def\a{\alpha}
\section{Introduction}
Over the few last years there has been a great interest in studying the embedding of supersymmetric QFT in string compactification on Calabi Yau manifolds using geometric engineering method. In this approach, one gets many precious informations on the non perturbative dynamics of string low energy QFT just by help of general results on toric geometry and local mirror symmetry.  For example, 10d type IIA string compactification on the $A_{n-1}$ local geometry develops an enhanced  $SU(n)$ gauge symmetry in six dimensions and a further compactification on the complex curve $P^1$ leads to a  $N=2$  QFT in four dimensions \cite{klmvw,kkv}. 10d type IIA compactifications on Calabi Yau threefolds which has ADE singularities over a complex one dimensional base space B with no one cycles lead to a pure 4d $N=2$ supersymmetric gauge theory whose gauge groups has simply laced Lie algebras of ADE type \cite{kmv}. Matters come form considering extra singularities located above special points of B  \cite{kmv,m, bfs1,bfs2,sv,tv,b}.\\
 Geometric engineering of supersymmetric (conformal)  QFT can be extended to include more general gauge groups having non simply laced Lie algebras such as the  $SO(2n+1)$ and $SP(n)$ gauge groups. In this regards, it was suggested in \cite{kmv,m} that the analysis developed for the simply laced gauge symmetries may be extended to the non simply laced ones by using the well known techniques of folding the nodes, of the ADE Dynking diagrams, permuted by outer auotmorphisms. For example, the non simply laced groups $SO(2n+1)$ and $SP(n)$ may be respectively obtained from $SO(2n+2)$ and $SU(2n)$ by help of the $Z_2$ outer automorphisms of the $D_n$ and $A_n$ Dynkin diagrams. The $Z_2$ symmetry exchanges the two end nodes of the $D_n$ Dynking diagram and acts as a reflection fixing one node in the $A_n$ Dynking diagram. A program of engineering all supersymmetric QFTs embedded in string and F-theory compactifications in presence of D branes has been initiated in \cite{klmvw, kkv, kmv}. Since then several results, including exact ones, were obtained for type II strings on Calabi-Yau manifolds and the main lines for the low energy F-theory non perturbative QFT dynamics were drawn \cite{vf,bm,m1,m2,v1}.\\
The aim of this paper is to contribute to this program by completing some results on the study of supersuymmetric QFTs with non simply laced gauge groups embedded in superstring and F-theory compactifications on Calabi-Yau manifolds \cite{vf,vm}. In particular we will discuss the following four:\\
(1) complete the analysis of \cite{bm,m1,m2} by showing that, in general, there exists two classes of solutions depending on two possible realisations of the outer-automorphism goups of $\Delta$(ADE). Each of these representations leads to a precise class of solutions for the mirror potential. The first class is the known one, while the second one is new.\\
(2) develop the study of F-theory compactifications on elliptically fibered Calabi-Yau considered recently in \cite{m1,v1}. We give explicit results for affine non simply laced toric data and the corresponding mirror geometries extending known results obtained in literature for the affine ADE cases.\\
(3) complete the analysis for ordianry ADE singularities by giving the explict derivation of the lacking non simply ones in type II superstring compactifications on local Calabi-Yau threefolds. Here also we give two kinds of solutions for the mirror potentials.\\
(4) finally, we initiate the basis of polyvalent toric geometry extending bivalent and trivalent geometries considered in geometric engineering method of supersymmetric QFTs. As an application, we use this extended toric method to derive a new solution for the affine $ D_4$ singularity. Recall that the affine $ D_4$ Dynkin diagram corresponds to a special geometry as it has a tetravalent vertex; i.e, a central vertex from which emanate four legs. This particular geometry is not recovered in the classification made in \cite{kmv}; though a special solution has been considered recently in \cite{bfs1}. Here, we will reconsider this question and complete the obovementioned results.\\
The presentation of this paper is as follows: In section 2, we review the main lines of toric method for studing local Calabi-Yau manifolds and give some general results. In section 3,  we give the machinery used in the study of the resolution of ADE singularities of local K3 surfaces, which we illustrate on the example of $A_{n-1}$ space. In section 4, we introduce the basic idea of geometric engineering of supersymmetric QFTs embedded in string theory. First we study the $N=2$ QFTs in terms of D branes of type II strings compactifications on singular Calabi-Yau manifolds. Second, we consider F-theory compactifications to engineer $N=1$ gauge theories in eight dimensions with ADE gauge groups. In section 5, we derive the toric realizations of folding under the auter-automorphisms of affine ADE Cartan graphs. Then, we extend the known results obtained in literature for affine ADE models to the non simply laced affine BCFG ones. In section 6, we use the techniques developed in section 5 to derive a new solution for the mirror potential associated to supersymmetric QFTs embedded in type II strings on Calabi-Yau threefolds with ordinary BCFG singularities.
 In section 7, we develop polyvalent geometry for higher dimensional toric manifolds, which may be viewed as an extension of the bivalent and trivalent geometries, and explore the leading tetravalent term. In this case, we discuss two constructions for affine so(8) geometry. The first one involves a rational toric lattice and leads to double valued mirror potential for Calabi-Yau twofolds while the second concerns Calabi-Yau threefolds compactifications and involves an integer toric lattice leading to single valued mirror potentials. In section 8 we give our conclusion.

\section{  Generalities on Toric geometry}
Here we review briefly some basic facts about toric geometry useful for our later analysis of geometric engeeniring of non perturbative supersymmetric quantum field theories embedded in superstrings and F- theory compactifications on Calabi-Yau manifolds. To start, let us recall that toric geometry is a tricky method for describing the essential one needs about the $n$-dimensional toric manifolds involved in superstring and F-theory compactifications. Roughly speaking, toric manifolds are complex $n$-dimensional manifolds exhibiting toric actions $U(1)^n$ which allow to encode the geometric properties of the complex spaces in terms of simple combinatoirial data of polytopes ${\Delta}_n$ of the $R^n$ space \cite{f,c,ba,lv,cpr,r,s}. In this correspondence, fixed points of the toric actions $U(1)^n$ are associated with the vertices of the polytope ${\Delta}_n$, the edges are fixed one dimensional lines of a subgroup $U(1)^{n-1}$ of the toric action $U(1)^n$ and so on. The beauty of this representation is that it permits to learn the essential about the geometric features of toric manifold by simply knowing the data of the corresponding polytope ${\Delta}_n$. To illustrate the main idea of this method, let us describe succintly the philosophy of toric geometry by giving two examples;  then we turn to describe with some mathematical details how things work in practice .\\
 {\it (i) $P^1$ complex curve}.\\
 This is the simplest example in toric geometry which turns out to play a crucial role in our analysis; first because it is a building block of higher dimensional toric manifolds and second because of the role it plays in the study of the blow up of singularities of Calabi-Yau manifolds involved in superstring compactifications. $P^1$ has a $U(1)$ action having two fixed points P and Q, generally chosen as $P=-1 $ and $Q=1$, describing respectively north and south poles of the real two sphere $S^2 \sim  P^1$. The correponding one dimensional polytope is just the straight line $[P,Q]$ joing the two points P and Q. \\
{\it (ii) $P^2$ projective space.}\\
 This is the second simplest example we want to give; it can be extended immediately to the higher dimensional complex projective space $P^n$, ( $n \geq 3$). There different ways to introduce the $P^2$ toric manifold; an interesting one is through embedding in $C^3$. $P^2$ is a complex two dimensional manifold with a $U(1)^2$ toric action exhibiting three fixed points $V_1$, $V_2$ and $V_3$. The corresponding polytope ${\Delta}_2$ is a finite sublattice of the $Z^2$ square lattice; it describes the intersection of three $P^1$ curves defining a triangle ($V_1 V_2 V_3$) in the $ R^2$ plane. A convenient choice of the data of the three vertices are: $ V_1= (1,0)$, $V_2=(0,1)$, and $ V_3=(-1,-1)$. Note that ${\Delta}_2$ has three edges namely $[V_1,V_2]$, $[V_2,V_3]$ and $[V_3,V_1]$ stable under the three $U(1)$ subgroups of $U(1)^2$; two subgroups are just the two $U(1)$ factors while the third subgroup is the diagonal one.\\
\subsection{Complex projective spaces}
Here we want to describe briefly how the toric geometry machinery works for complex projective spaces $ P^n $. This analysis is also valid for weighted $n$-dimesional projective spaces $ WP^n $ and complex hypersurfaces in $ P^n $ and $ WP^n $. In the $C^{n+1}$ space parametrised by the local coordinates $(z_1,\ldots,  z_{n+1})$, the $ P^n $ manifold can be expressed as:
\begin {equation}
P^n= {C^{n+1}-\{(0,\ldots,0)\}\over C^*},
\end{equation}
 where  the $ C^*$ operation acts as:
   \begin {equation}
 (z_1,\ldots,  z_{n+1}) \to ( \lambda z_1,\ldots,\lambda z_{n+1}).
\end{equation}
In the language of toric geometry, we associate to $ P^n $ a $ n$-dimensional toric diagram ${\Delta}_n({P_n})$ ( $n$-dimensional simplex ) with $(n+1)$ vertices $v_i$; $i=1,\ldots, n+1$. The relation(2.2) is translated to a condition on the vertices $v_i$ which should add to zero.
\begin {equation}
  v_1 + \ldots + \ldots v_{n+1} =0.
\end{equation}
The solution of this eq is given by $v_{i}= e_{i}$, $i=1,\ldots,n$ and $v_{n+1}= - \sum\limits_{i=1}^ne_i$, where the $e_{i}$'s are the $Z^n$  canonical basis.
Note that the above eq is in fact a special relation among the vertices which may be extended naturally as:
\begin {equation}
  k_1v_1+ \ldots +k_{n+1}v_{n+1} =0.
\end{equation}
where the $ k_i's ; i=1,\ldots, n+1$ are arbitrary integers. The $ n$-dimensional simplex satisfying such a relation defines just the toric diagram ${\Delta}_n({WP_n})$ of the weigthed projective space $ (WP^n = \{ z_i=\lambda^{k_i}z_i,\quad  i=1,\quad n+1 \}$ with    $(k _1,  \ldots k_{n+1} )\neq (1,\ldots,1)$).
\subsection{Toric manifold $M_{\Delta}^n$}
The above toric construction of the projective spaces can be extended to more general $n$-dimensional toric manifolds $M_{\Delta}^n$ embedded in $ C^{k}-U$, $n=k-r$; where now we have a ${C^*}^r$ action, $r>1$  and instead of substracting the origin $((0,\ldots,0)$ of $C^{k}$, we now substruct a non trivial open set U of $C^{k}$. In other words, $M_{\Delta}^n$ is defined as:
\begin {equation}
M_{\Delta}^n = {C^{k}-U\over {C^*}^r},
\end{equation}
where the ${C^*}^r$ action acts as
\begin {equation}
 {C^*}^r: z_i   \to \lambda ^{q_i^a} z_i ,\quad  i=1,2,\ldots ,k ; a=1, 2,\ldots ,r
\end{equation}
 and where the U open set is given by:
\begin {equation}
 U= \cup _I \{ (z_1,\ldots, z_{k})| z_i=0 \quad   for  \, all \; i  \in I \}.
\end{equation}
  Like for complex projective spaces, the toric manifold $M_{\Delta}^n$ can be encoded in toric diagram $ \Delta(M)$ having $ k=n+r$  vertices $ v_i$ generating a $n=k-r$ dimensional sulattice of the  $Z^d$ lattice and satisfying the following $r$ relations \cite{kmv},
    \begin {equation}
  \sum \limits _{i=1}^{n+r} q_i^a v_ i=0,\quad a=1,\ldots,r.
\end{equation}
The above toric analysis we have been describing, and which may be also extended to fiber bundles of toric manifolds, has an interesting physical interpretation in terms of vaccua solutions of $2d$  $N=2$ supersymmetric linear sigma model \cite{w,agm} and superstring Calabi-Yau compactifications. Indeed interpreting the previous $z_i$ coordinates as $(\phi_i)$ matter fields and the $ q_i^a $ integers as the quantum charges the $(\phi_i)$'s under a $ U(1)^r$ symmetry, then toric manifolds are just the target space of the $2d$ $N=2$ supersymmetric linear sigma model. The $ q_i^a $'s obey naturally the neutrality condition which means that the theory flows in the infrared to a non trivial superconformal model \cite{kmv,bs1,bs2}. The vaccum energy of such $N=2$ linear sigma model is given by the so called D-flatness eq namely;
     \begin {equation}
  \sum \limits _{i=1}^{d+r} q_i^a |\phi _i|^2=R_a.
\end{equation}
In this eq, the $ R_a$'s are Fayet-Iliopoulos (FI) coupling parameters which are interpreted as Khaler moduli of the manifold. We have different cases to consider here:\\
 (i) If the  $ q_i^a $'s are all positive definite, or negative definite, the toric target space is compact. However, if there is a mixture of positive and negative $ q_i^a $'s, the toric target space is non compact.\\
 (ii) If all $R_a$'s are zero, then the toric manifold is singular.\\
(iii) For all $R_a$'s $\neq 0$, we have a smooth toric manifold. In this case the FI parameters are interpreted as blow up parameters of the singularity of the toric manifold.
 \subsection{Calabi-Yau varieties as toric manifolds}
In superstring compactifications, one is generally interested in complex $d$-dimensional Calabi-Yau manifolds. These are Ricci flat manifolds with a vanishing first Chern class and a holonomy contained in the unitary $SU(d)$ symmetry. A large class of Calabi-Yau manifolds can be realized as hypersurfaces in toric varieties described previously. Commun examples of $d$-dimensional Calabi-Yau manifolds are described by hypersurfaces in the weighted projective space $ WP^{d+1} (k _1, k_2, \ldots k_{d+2} )$, with the identification (2.6). Put differently, a d-dimensional Calabi-Yau manifold is represented by a quasi-homogenous polynomial $P{_\delta}(z_1,\ldots,z_{d+2})$ of degre $ \delta$. Its general form is given by \cite{ba,lv}:
  \begin {equation}
  P_{\delta}(z_1,\ldots,z_{d+2})=\sum \limits _{\mu{_1},\ldots,\mu_{d+2}}{ a _{ \mu{_1},\ldots, \mu_{d+2}}{ z{_1}^{\mu_1}\ldots z_{d+2}^{\mu_{d+2}}}}.
\end{equation}
In this eq $(\mu{_1},\ldots,\mu_{d+2})$ are integers, which upon taking the scalar product with the vector $(k _1, k_2, \ldots k_{d+2} )$, define the degre $\delta$ of the monomials of $P_{\delta}$; i.e $\delta={\vec\mu}\vec k$. Note that the previous potential may be rewritten by using the (d+1) gauge invariants ($ Z_1,\ldots,Z_{d+1}$) under the toric action. Straightforward calculations shows that eq(2.10) can be put as follows:
\be
\sum\limits_{{\vec r}\in\Delta}{ b _{ r{_1},\ldots, r_{d+1}}{ Z{_1}^{r_1}\ldots Z_{d+1}^{r_{d+1}}}}=0,
\ee
where $(r_1,\ldots,r_{d+1})$ are vertices of the polytope of $WP^{d+1}$.
\subsection{ Mirror manifolds }
Given a toric manifold $M_d$, one can build its dual manifold $ W_d$, currently known as the mirror manifold. This is also a toric variety which is obtainaid from $M_d$ by help of mirror symmetry \cite{ba,lv,fhsv,ckyz,hv,hiv}. The latter is a symmetry which associates to any $d$- dimensional Calabi-Yau $M_d$ of Hodge numbers $h^{1,1}(M)$ and $h^{d-1,1}(M)$, a mirror manifold $W_d$ whose Hodge numbers $h^{1,1}(W)$ and $h^{d-1,1}(W)$ are constrained as:
   \begin {equation}
 \begin{array}{lcr}
h^{1,1}(M)= h^{d-1,1}(W)\\
h^{d-1,1}(M)=h^{1,1}(W).
 \end{array}
\end{equation}
 This means that the complex (Kahler) moduli space of $M_d$ is identical to the Khaler (complex) moduli space of $W_d$. Mirror symmetry plays an important role in superstring compactifications, string dualities and in the geometric engineering of supersymmetric quantum field theory where it can be naively viewed as a generalized version of T duality \cite{syz}.\\
 In the toric representation of the local Calabi-Yau manifolds we are interested in here, mirror symmetry maps the toric relations between the vertices (2.8) of $M_d$ to relations between the monomials $y{_i}= \prod \limits _j\tilde z_j^{<{\tilde v}_j, v_i>}$ in the defining equation of the mirror manifold $W_d$. The $\tilde z_j$'s and  ${\tilde v}_j$ are respictively the dual variables  and vertices of $W_d$. The mirror geometry is then given by:
\begin {equation}
  \sum \limits _i {a{_i}y_i}=0,
\end{equation}
where the $ a_i$'s are complex parameters defining the complex structure of $W_d$ and $y_i$ are gauge invariant monomials satisfying the following remakable identity:
  \begin {equation}
   \prod \limits _i y_i^{q^a_i}=1.
\end{equation}
 Note that the solution of these equations represent non compact hypersurfaces with $d-2$ dimensions and seems to not reproduce the right dimension. This is not really a problem since we can usually restore the correct dimension by using the standard trick which consists to introduce two auxiliary variables $u$ and $v$ and rewrite eq(2.14)as \cite{kmv}:
  \begin {equation}
  \sum \limits _i a_iy_i=uv.
\end{equation}
This hypersurface has now the right dimension; the quadratic term $uv$ involved in this equation does not affect the complex stucutre of the mirror geometry. Having introduced compact, non compact and singular toric manifolds as well as their mirrors, we turn now to explore further their singularities.
\section{ ADE surfaces}
ADE geometries are singular complex two dimensional surfaces which play a cental role in the investigation of the non perturbative dynamics of the quantum field theory limit of superstring compactifications on ALE surfaces. The resolution of the singularity of these surfaces is nicely described in toric geometry; the corresponding toric polytopes $\Delta$(ADE) have a very remarkable structure as they look like the usual Dynking diagrams of ordinary and affine ADE Lie algebras. The singularities of these ALE spaces are then of two types:\\ (i) Ordinary singularities classified by the ordinary ADE Lie algebras.\\ (ii) Affine singularities classified by the Affine ADE Kac-Moody algebras. The latters are at the basis the derivation of the new $4d$ $N=2$ superconformal field theories \cite{kmv,bfs2}.\\
Before going ahead it should be noted that one may also consider ALE spaces with BCFG singularities. We will see later that the study of these surfaces differs a little bit from the study of the ADE ones. The potentials defining these complex surfaces are generally non single valued functions and deserves a careful analysis. For the moment, let us focus our attention on reviewing briefly the main lines of the ADE geometries. \\
In $C^3$ parametrised by the local coordinates $x,y,z$, ordinary ADE geometries are described by the following complex surfaces \cite{hov}.
\begin {equation}
\begin{array}{lcr}
A_n: xy+z^n=0\\
D_n: x^2+y^2z+z^{n-1}=0\\
E_6: x^2+y^3+z^4=0\\
E_7: x^2+y^3+yz^3=0\\
E_8: x^2+y^3+z^5=0.\\
\end{array}
\end{equation}
These equations have singularities located at the $C^3$ origin $ x=y=z=0$ and are resolved in two ways; either by deforming the complex structure of the surface or varying its Kahler structure. The two deformations are equivalent due to the self mirror property of the ALE space. This is why we shall restrict ourselves hereafter in giving only the kahler deformation.\\
\\
{\bf  Kahler deformation}\\
 This deformation consists on blowing up the singularity by a collection of intersecting complex curves; i.e replacing the singular point $(x,y,z)=(0,0,0)$ by a set of intersecting complex curves so that the resulting surface is no longer singular. The nature of the set of intersecting $P^1$ curves and the toric graph one obtains depend on the type of the singular surface one is considering. In sum, the toric polytopes $\Delta$(ADE) of the smoothed ADE surface share several features with the ADE Dynkin diagrams; in particular the intersection matrix  $q^a_i$(ADE) (known also as Mori vectors in toric language) of the complex curves used for the resolution of ADE singularities is, up to some details, the opposite of the ADE Cartan matrix $K_{ij}$(ADE). This link leads to a nice correspondence between the ADE roots $ \alpha$  and  $C_2$ two-cycles involved in the deformation of $ADE$ singularities. More specifically, to each simple root $ \alpha_i$, we associate a single real 2 sphere $(S^2)_i$.
This nice connection between toric geometry of the ADE surfaces and Lie algebra turns out to be at the basis of important developments both in physics and mathematics. In string theory, the abovementioned link has been successfully used in different occasions; in particular in: (i) The geometric engineering of the 4d $ N=2$ supersymmetric quantum field theory embedded in type II strings compatification in presence of D branes \cite{kmv}.(ii)Geometric engineering of $ N=1$ models from F-theory compactification \cite{v1}.(iii) The development of a new method for resolving singularities using D branes physics. We shall turn to these relations in more details in section 4.\\
 In what follows, we want to use this correspondence to explore further the singularities of the ALE surfaces and their resolutions. To that purpose, we first consider the ordinary $A_{n-1}$ example by collecting the following useful informations:\\
(i) the simple roots $\alpha_i$ in terms of a real basis $\{e_i\}$ of the orthonormal vectors  of $R^n$.\\
(ii) the fundamental weights, $\lambda_j$, in terms of the simple roots and the $\a$'s.\\
 (iii) the adjoint representations $\lambda_{adj}$ and the outer-automorphism groups, $\Gamma$(ADE), of the Dynkin diagram.\\
 We describe the $A_{n-1}$ case in some details and give later the results for the other simply laced Lie algebras; they may be obtained without difficulty.
 \subsection {$A_{n-1}$ hypersurfaces  }
We begin by giving the above-mentioned materials, then we make some comments regarding the link with Kahler and complex resolutions of the  $A_{n-1}$ singularity of the ALE space.\\
 \qquad {Simple roots}:
\begin{equation}
\alpha_j=e_j-e_{j+1}, j=1,...,n-1.
\end {equation}
\qquad { Fundamental weights }$\lambda_k(A_{n-1})$
\begin{equation}
\begin{array}{lcr}
 \lambda_k&=\{{1\over n} {\sum\limits_{j=1}^{k-1}(n-k)j\alpha_j+\sum\limits_{j=k}^{n-1}k(n-j)\alpha_j}\},\quad k=1,\ldots,n-1\\
 &=\{{1\over n} {\sum\limits_{j=1}^{k}(n-k)je_j-\sum\limits_{j=k+1}^{n}ke_j}\}.
\end{array}
\end {equation}
\qquad {  Adjoint representation}:$\lambda_{adj}(A_{n-1})$
\begin{equation}
\begin{array}{lcr}
 \lambda_{adj}=& \lambda_1+ \lambda_{n-1}&(a)\\
\lambda_{adj}=&e_1-e_n&(b)\\
\lambda_{adj}=&\sum\limits_{j=1}^{n-1}\alpha_j&(c).
\end{array}
\end {equation}
We shall see later on that the eq(3.4-b) define the $A_{n-1}$ singularity while eq(3.4-c) describes its Kahler resolution or equivalently the complex resolution by using mirror symmetry \cite{kmv}.\\
\qquad {  Cartan matrix }$K(A_{n-1} )$
\begin{equation}
\begin{array}{lcr}
 K_{ij}=\alpha_i\alpha_j,\quad 1\geq i,j\leq n-1\\
=(e_i-e_{i+1}).(e_j-e_{j+1})\\
=2\delta_{i,j}-\delta_{i,j-1}-\delta_{i,j+1}.
\end{array}
\end {equation}
Observe here that $\sum\limits_{j=1}^{n-1}K_{ij}$ vanishes for all values of $i$ between 2 and $n-2$. For $i=1$ and $n-1$,$\sum\limits_{j=1}^{n-1}K_{ij}=1$. Thus, we have:     \begin{equation}
\sum\limits_{j=1}^{n-1}K_{ij}=\delta_{i,1}+\delta_{i,n-1}.
\end {equation}
In toric geometry, this equation is interpreted as the Calabi-Yau condition of the blown up $A_{n-1}$ surface.\\
 { Outer automorphism group $\Gamma (A_{n-1})$}\\
Recall that the outer automorphism group $\Gamma (g)$ of a Lie algebra g is a discrete group leaving the Dynkin diagram of g invariant. In case of $(g=A_{n-1})$,  $\Gamma (A_{n-1})$ is:
 \begin{equation}
\Gamma={Z\over{2Z}}\approx Z_2.
\end {equation}
In the $ A{_{2k}-1}$  $(n=2k)$, $k \geq 2$ case, $ \Gamma$ acts non trivially on $2k-2$ nodes of the diagram. It leaves the middle node, associated with the simple root $\alpha_k$, stable and permutes the $2k-2$ remaining ones as:
\begin{equation}
\begin{array}{lcr}
\alpha_{k+j}\leftrightarrow \alpha_{k-j},j=1,...,k-1\\
e_{k+j}\leftrightarrow e_{k-j}.
\end{array}
\end {equation}
The above algebraic data translates to the language of toric geometry as follows. (i) The $e_i$  basis vectors of $R^n$ are associated with the fix points of the toric action of a complex one dimensional 2-cycle $C^2$ having no one cycles. In other words, the blow up of the $A_{n-1}$ geometry is a collection of $(n-1)$ intersecting compact $P^1$ curves. (ii) The entries $ q_i^a$ of the Mori vectors $ q^a ,a=1,...,n-1$, describing the intersections of the 2- cycle are divided in two blocks: First the block given by the following $(n-1)\times(n-1)$ square matrix.
\begin{equation}
q_i^a=-K_{ia},\quad  i=1,...,n-1.
\end {equation}
The remaining vectors $q_0^a$ and $q_n^a$ are obtained by introducing the following imaginary simple roots $\alpha_0^*  $ and $\alpha_n^*  $  defined as
\begin{equation}
 \alpha_0^*=  e_0^*-e_1 , \quad  \alpha_n^* =e_n-e_{n+1}^*  ,
\end {equation}
where  $e_0^*$ and $ e_{n+1}^*$ are such that
 \begin{equation}
\begin{array}{lcr}
{ e_0^*}^2= -1,\quad   {e_{n+1}^*}^2=-1\\
e^*_0 e^*_{n+1}=e^*_0 e_i=e^*_{n+1 }e_i=0, i=1,...,n.
\end {array}
\end {equation}
The extra vectors $e^*_0$  and  $e^*_{n+1 } $  may be viewed as basis vectors obtained from the complexification of the $e_1$  and $e_n$  axes respectively that is  $e^*_0=ie_1'$, $e^*_{n+1 }=ie'_n$. They represent fix points located at infinity of the toric action of two non compact spheres $e^*_0-e_1$ and  $e_n-e^*_{n+1 }$. This procedure allows us to recover the extra dimension of the $A_{n-1}$ geometry of the ALE space as well as the Calabi-Yau condition,         \begin{equation}
\sum\limits _{i=0}^n q_i^a=0,
\end {equation}
where now the index $i$ runs over all vertives of the toric polytope of the  $ A_{n-1}$ geometry.
Taking into account eqs (3.10-3.11), the modified Cartan matrix $\tilde K_{ij}$  compatible with the geometry of the ALE space may defined as
\begin{equation}
\begin{array}{lcr}
 \tilde K_{ij}=K_{ij}\quad i,j=1,\ldots,n\\
\tilde K_{0j}=\tilde K_{nj}=0,\quad j=0,\ldots,n \\
\tilde K_{i0}=-\delta_{i,1}\\
\tilde K_{ij}=-\delta_{i,n-1}.
\end {array}
\end {equation}
Since the $ i=0 $ and $i=n$ rows of  $\tilde K_{ij}$ are identically zero, one may replace this $ (n+1)\times (n+1)$ square matrix by a $ (n-1)\times( n+1)$ rectangular matrix  $K_{aj} = -q_j^a ,a=1, \ldots, n-1, j=1,\ldots, n-1$ given by
 \begin{equation}
q_j^a=\delta^a_{j-1}-2\delta^a_{j}+\delta^a_{j+1}.
\end{equation}
Eqs (3.14) give the entries of the Mori vectors satisfying automatically the Calabi Yau condition (3.12). The dimension of the ALE space is just $(n+1)-(n-1) = 2$.  In toric geometry language, the toric diagram of the blowing up the  $A_{n-1}$ is given by $( n+1)$ vertices $ v_i$  of the standard  $Z^2$
 lattice such that
\begin{equation}
\begin{array}{lcr}
\sum \limits _{i=0}^{n} q^a_iv_i=0 \\
\sum \limits _{i=0}^{n} q^a_i=0, a=1,2,\ldots,n-1
\end{array}
\end{equation}
\subsection {$D_{n}$ and $E_{r}$ surfaces}
{\bf  \it 1-  Ordinary $D_{n}$ algebras}\\
This is the algebras of the $ SO(2n)$ orthogonal group.\\
 \qquad { Simple roots}:\\
\bea
\alpha_j&=&e_j-e_{j+1}, \quad j=1,...,n-1\nn\\
\alpha_n&=&e_{n-1}+e_n
\eea
\qquad { Fundamental weights }$\lambda_k(D_{n})$
\bea
 \lambda_k&=\{{1\over n} {\sum\limits_{j=1}^{k-1}j\alpha_j+\sum\limits_{j=k}^{n-1}\alpha_j} +{ k \over 2}( \alpha _{n-1}+\alpha_n)\}=\lambda_k&=\sum\limits_{j=1}^{k}e_j,\quad k=1,\ldots,n-2.
\eea
\qquad {  Adjoint representation}:$\lambda_{adj}(D_{n})$
\begin{equation}
\begin{array}{lcr}
 \lambda_{adj}=& \lambda_2&(a)\\
\lambda_{adj}=&e_1+e_2&(b)\\
\lambda_{adj}=&\alpha_1+\sum\limits_{j=1}^{n-1}\alpha_j+(\alpha_{n-1}+ \alpha_n)&(c).
\end{array}
\end {equation}
\qquad {  Cartan matrix }$K(D_{n} )$
\begin{equation}
\begin{array}{lcr}
 K_{ij}=K_{ij}(A_{n-2},\quad  ;i,j=1,\ldots,n-2\\
=(e_i-e_{i+1}).(e_j-e_{j+1})\\
=2\delta_{i,j}-\delta_{i,j-1}-\delta_{i,j+1}.
\end{array}
\end {equation}
 { Outer automorphism groups $\Gamma (D_{n})$}\\
 \bea
\Gamma (D_n)=Z_2:\quad
\alpha_{n-1}&\leftrightarrow&\alpha_{n},\quad n\geq 5 \nn\\
\alpha_{i}&\leftrightarrow& \alpha_{i},\quad i=1,...,n-2\nn\\
\eea
\begin{equation}
\begin{array}{lcr}
 \Gamma (D_4)={\cal S}_3:\\
(\alpha_{1},\alpha_{3}\alpha_{4})&\leftrightarrow&
(\alpha_{i1},\alpha_{i2}\alpha_{i3})\\
\alpha_{2}&\leftrightarrow&\alpha_{2}
\end{array}
\end {equation}
2-{\bf  Ordinary $E_n$ algebras}\\
This is a continuation of the previous algebras where we summarize the main algebras features of  these excptional Lie algebras.\\
Ordinary $E_6$ algebras\\
 \qquad { Simple roots}:
\bea
\alpha_1=e_1-e_2, \alpha_2=e_2-e_3,\alpha_3=e_3-e_4\nn\\
\alpha_4=e_4-e_5, \alpha_5=e_1+e_2,\\
\alpha_6= {1\over 2}(e_1+(e_6-e_2 + \ldots +e_5))\nn
\eea
 \qquad {  Adjoint representation}:$\lambda_{adj}(E_{6})$
\begin{equation}
\begin{array}{lcr}
\lambda_{adj}=\alpha_1+ \alpha_5+2(\alpha_2+\alpha_4+\alpha_6)+3\alpha_3.
\end{array}
\end {equation}
{ Outer automorphism groups $\Gamma (E_{6})$}\\
 \bea
\Gamma (E_6)= Z_2: \quad
\alpha_{i}&\leftrightarrow & \alpha_{6-i},\quad i=1,2,3\nn\\
\alpha_{6}&\leftrightarrow&  \alpha_{6}.
\eea
Ordinary $E_7$ algebras\\
 \qquad { Simple roots}:\\
\bea
\alpha_1=e_1-e_2, \alpha_2=e_2-e_3,\alpha_3=e_3-e_4\nn\\
\alpha_4=e_4-e_5, \alpha_5=e_5-e_6, \alpha_6=e_1+e_2,\\
\alpha_7= {1\over 2}(e_1+e_7-(e_2 + \ldots +e_6)),\nn\\
\eea
 \qquad {  Adjoint representation}:$\lambda_{adj}(E_{7})$
\begin{equation}
\begin{array}{lcr}
\lambda_{adj}=2\alpha_1+3\alpha_2+4\alpha_3+3\alpha_4+2\alpha_5 +\alpha_6+2\alpha_7.
\end{array}
\end {equation}
{ Outer automorphism groups $\Gamma (E_{7})$}\\
 \bea
\Gamma (E_7)=1: \quad
\alpha_{i}&\leftrightarrow & \alpha_{i},\quad i=1,\ldots,7
\eea
Ordinary $E_8$ algebras\\
 \qquad { Simple roots}:\\
\bea
\alpha_1=e_1-e_2, \alpha_2=e_2-e_3,\alpha_3=e_3-e_4\nn\\
\alpha_4=e_4-e_5, \alpha_5=e_5-e_6, \alpha_6=e_6-e_7,\\ \alpha_7=e_1+e_2,\alpha_8= {1\over 2}(e_1+e_8-(e_2 + \ldots +e_7)),\nn\\
 \eea
 \qquad {  Adjoint representation}:$\lambda_{adj}(E_{8})$
\begin{equation}
\lambda_{adj}=2\alpha_1+2\alpha_2+3\alpha_3+4\alpha_4+3\alpha_5 +\alpha_6+2\alpha_7.
\end {equation}
{ Outer automorphism groups $\Gamma (E_{8})$}\\
\bea
\Gamma (E_8)=1: \quad
\alpha_{i}&\leftrightarrow & \alpha_{i},\quad i=1,\ldots,8
\eea
 In the end of this section, it is interesting to note that the toric method used for ADE spaces is also valid for the ordinary and affine non simply laced BCFG ones. We will propose later on some tools useful for the study of non simply laced mirror geometries. For the moment let us expose how toric methods are exploited in geometric engineering of QFT.
\section{ Toric Geometry in superstring compactifications. }
\subsection{$4d$ $N=2 $ QFT from type II superstrings}
It is now quite well estabished that there are essentially two methods to study non perturbative dynamics of supersymmetric quantum field theories one gets from the low energy limit of superstring theories. (i) The first way is based on D-p brane configurations on which leave (p+1)-dimensional gauge theories. This manner of doing has been extensively studied in the physical litterature \cite{hw,wbrane}and knows nowadays a great interest of attention in connection with noncommutative geometry \cite{swn}; in particular with noncommutative euclidean four dimensional instantons \cite{ns,bhss}, noncommutative solitons \cite{gn}and tachyon condensation in open string field theory \cite{a}. (ii) The second way, which we will consider here, is based on considering  supersymmetric quanum field theories (QFT) realized as singular limits of superstring and F- theories on local $K3$ fibration Calabi-Yau manifolds M on a base B. In this geometric construction one mainly use toric geometry tools and local mirror symmetry we discussed earlier. A remarkable feature of this representation is that the non perturbative gauge group G of the QFT is obtained from the singularities of the fiber $K3$ and matter is given by non trivial geometry on the base B. The space of physical parameters of the QFT is related to the moduli space of both the fiber F=K3 and the base B  of $M$; in particular the gauge coupling $g$ is proprortional to the inverse of the square root of the volume  $ V(B)$of the base; i.e:
\be
V(B)=g^{-2}.
\ee
\subsubsection{More on Geometric engeneering method}
To start let us recall that among the important consequences of superstrings dualities is that many non trivial facts for supersymmetric Quantum field theories have found natural explanation in the context singular limit of superstrings on local Calabi-Yau manifolds. Several exact results for the Coulomb branch of $4d$ $ N=2 $  QFT, generalizing the Seiberg -Witten (SW) model \cite{swg1,swg2}, are naturally obtained by help of local mirror symmetry of type II superstring compactifications on Calabi-Yau threefolds $M_3$. The latter is realized as a K3 fibration with ADE singularity on a $P^1$ complex curve or a collection of intersecting $P^1$ curves. Moreover duality between heterotic superstring on $ K3\times T^2$ and type IIA superstring on $ M_3$, shows that the relevant part of the QFT moduli space comes from the ADE type singularities of $M_3$ \cite{ketal,kv}. This observation permitted to: (i) rederive the known exact results for the $4d$ $ N=2$ Seiberg-Witten model. (ii) observe that the 4d gauge fields of the $N=2$ supersymmetric QFT are just the gauge fields one gets from type IIA  D2-branes wrapping over $K3$ vanishing cycles. (iii) matter is obtained from extra singularities in the base B of $M_3$.
This idea was first considered in \cite{klmvw}and was further developed in \cite{kkv}and was called geometric engineering of $4d$ $ N=2$ QFT.
\subsubsection{$N=2$ ADE model from type II superstrings }
The main steps in getting $4d$ $N=2 $ QFT from type IIA superstring on Calabi Yau threefolds may be summarized as follows: First specify the type of ADE singularity of the $K3$ fiber. Then consider the limit where the volume $V(B)$ of the base of $M_3$ is very large so that gravitational effects may be ignored. Finally examine the propagation of the type IIA on $M_3$ in presence of D2 branes wrapping on 2-cycles of K3. To illustrate the method, suppose that $K3$ has a $su(2)$ singularity. In the vicinity of the $su(2)$ singularity, the fiber $K3$ may be viewed as an ALE space with $A_1$ singularity described by the following equation \cite{kmv,bfs2}:
$$ xy=z^2,$$
where $x, y$ and $z$ are complex variables. Taking the $x, y$ and $z $ coordinates of the ALE space as:
\bea
 x=&\phi_1^2\phi_2\nn\\
 y=&\phi_3^2\phi_2\\
 z=&\phi_1\phi_2 \phi_3\nn,
\eea
where now the $\phi_i$'s should be though of as complex fields of two dimensional  $N=2$ supersymmetric $U(1)$ gauge model, one discovers that eq(2.9) is intimately related with D-term of the bosonic potentiel $U(\phi_1, \phi_2,\phi_3)$ of supersymmetric theories with four supercharges:
\be
U(\phi_1, \phi_2,\phi_3)=(\phi_1 \phi_1+\phi_3\phi_3-2\phi_2\phi_2-R)^2.
\ee

In this eq, $R$ is the coupling parameter of the $U(1)$ Fayet-Iliopoulos (FI)term one may add to the lagrangian model, which reads in the superfield language as:
\be
L(\Phi,V)= \int d^4\theta \bar\Phi e^V \Phi -R\int d^4\theta V,
\ee
where $\Phi$ and $V$ are respectively the chiral and gauge superfields.
 From this presentation, one sees that the $U(1)$ Cartan subgroup of the $SU(2)$ symmetry of the singularity of $ K3$ carries the gauge symmetry of the linear sigma model. The presence of the FI term resolves the singularity of the potential $U(\phi_1, \phi_2,\phi_3)$. Geometrically, this consists to replace the singular point $x=y=z=0$ by a $P^1$ curve parameterized by a new variable $ x'$ defined as $x'={x\over z}$. In the new local coordinates  $(x',y,z)$, the equation of the $A_1$ singularity may be rewritten as:
\be
  x' y=z;
\ee
which is no singular any more. In the field theory language, this corresponds to a positive value of the FI coupling R.
The next step is to consider the propagation of type IIA superstring in this background. In this case, D2-branes wrapping around the blow up $P^1$ curves $x'$ give two $W_\mu^{\pm}$ vector particles depending of the two possible orientations for wrapping. The particles have mass proportional to the volume of the blow up real 2-sphere $x' $. $W_\mu^{\pm}$ are charged under the $U(1)$ field $Z_0^\mu$ obtained by decomposing the type IIA superstring 3-form in terms of the harmonic form on the 2-sphere $x'$. In the limit where the blow up 2-sphere $x'$ shrinks, we get then three masslesss vector particles $W_\mu^{\pm}$  and $Z_0^\mu$ which form an $SU(2 )$ adjoint. We thus obtain a $N=2$ $SU(2)$ gauge symmetry in 6 dimensions. A further compactification on the base $B$ ,that is on a real 2-shpere, gives $N=2$ pure $SU(2)$ Yang-Mills in 4 dimensions. At this level we want to make two remarks: (i)The above geometric $SU(2)$ gauge theory analysis can be easily extended to all simply laced $ADE$ gauge groups and, up to some pertinent details, to non simply laced $BCFG$ groups as well. (ii) To incorporate matter, one should consider a non trivial geometry on the base B of $M_3$. For example, if we have a 2 dimensional locus with $SU( n)$ singularity and another locus with $SU( m)$ singularity and they meet at a point, the mixed wrapped 2 cycles will now lead to $(n ,m) $ $N=2$ bi-fundamental matter of the $SU(n)\times SU(m)$ gauge symmetry in four dimensions. Geometricaly, this means that the base geomerty of $M_3$ is given by two intersecting $ P^1$ curves whose volumes $V_1$ and $V_2$  define the gauge coupling constants $g_1$ and $g_2$ of the $SU(n)$ and $SU(m)$ gauge symmetries respectively. Fundamental matter is given by taking the limit $V_2$ to infinity or equivalently $g_2=0$ so that the $SU(m)$  group becomes now a flavor symmetry. Geometric engineering of the $4d $ $N=2$ QFT shows moreover that the analysis we have been describing recovers naturally some remarkable features which follows from the connection between toric geometry and Lie algebras. For instance, taking $m=n$ and identifying the $SU(m)$ gauge symmetry with $SU(n)$ by equating the $V_1$ and $V_2$ volumes, which imply in turn that $g_1=g_2$, the bifundamantal matter becomes then an adjoint one. This property is more transparent in the language of the representation theory; the adjoint of $SU(n+m)$ splits into $SU(n)\times SU(m)$ representations as:
\be
(n+m){\overline{(n+m)}}= n.{\bar n}+m.{\bar m} +{\bar n}.m+n.{\bar m},
\ee
 where $n.{\bar n}+m.{\bar m}$ gives the gauge fields and ${\bar n}.m+n.{\bar m} $ define the bifundamental matters. For more details see \cite{kmv}.
Geometric engineering of $ 4d $  $N=2$ QFT is really a tricky method to study $4d$ $ N=2$ QFT embedded in type IIA superstring theory on Calabi-Yau threefold. One the beauties of this method is that $4d$ $ N=2 $QFT's are represented by quiver diagrams where for each $SU$ gauge group factor one associates a node and for each pair of groups with bi-fundamental matter, the two corresponding nodes are connected with a line. These diagrams have a similar representation as the Dynkin diagrams of ordinary and affine simply laced Lie algebras. An other beauty of this formulation is that the developments in non perturbative supersymmetric QFT obtained over the few last years are nicely described in this approach. In this regards, it is worthwhile to mention the three following:\\
(i) The obtention of exact solutions of Coulomb branch of $4d$ $N=2$ QFT which are conveniently described in geometric engineering method using toric geometry and local mirror symmetry. The latter plays a crucial role as it maps type IIA superstring on $M_3$ into type IIB string on $W_3$. Under this symmetry, type IIA superstring on $M_3 $ is exchanged to type IIB superstring on $W_3$ where classical solutions of the Coulomb branch can be shown to be exact solutions.\\
 (ii)The obtention of new $4d$ $N=2$ superconformal theories and their classification in terms of affine ABCFG diagrams \cite{kmv,bfs2}. \\
(iii) The  gauge coupling moduli space of these theories is related to the moduli of flat connections on the torus. These moduli are interesting in the study of the duality between heterotic string and F-theory on elliptically fibered Calabi-Yau manifolds and also in geometric  construction of $ N=1$ QFT we are interested in here.
\subsection{$N=1$ ADE model from F-theory compactification }
\subsubsection{F-theory backgrounds}
 F-theory defines a non perturbative vacuum of type IIB superstring theory in which the dilaton and axion fields of the superstring are not constants. The latters introduce in the game an extra complex moduli which is interpreted as the complex parameter of an elliptic curve leading then to nonpertubative vacua of type IIB in a twelve dimensional space time \cite{vf}. F-theory may also defined by help of string dualities. As we will see in a moment F-theory on elliptically fibered Calabi-Yau manifolds may also be defined in terms of dual string models, but let us first review briefly some features of this theory. Recall that type IIB string model is a ten dimensional theory of closed superstrings with chiral $N=2$ supersymmetry. The bosonic fields of the corresponding low energy field theory are the graviton $g_{\mu\nu}$, the antisymmetric tensor $B_{\mu\nu}$ and the dilaton $\phi$ coming from the NS-NS sector and the axion $a$, the antisymmetric tensor fields $\tilde B_{\mu\nu}$ and the self dual four form $D_{\mu\nu\rho\sigma}$ coming from R-R sector. As we see, there is no non abelian gauge field in the massless spectrum of type IIB but instead contains D$p$-branes, with $p=-1,1,3,5,7$ and  $9$ on which live $A_{\mu}$ fields. Note that these extended objects are non perturbative solutions playing a crucial role in string dualities. Note also that type IIB theory has a non perturbative SL(2,Z) symmetry  for which the fields $g_{\mu\nu}$ and $D_{\mu\nu\rho\sigma}$ are invariant but the complex string coupling $\tau_{IIB}=a+ie^{-\phi}$ and the doublet $(B_{\mu\nu},\tilde B_{\mu\nu})$ of two forms are believed to transform as \cite{sch}:
\be
\tau_{IIB} \to {a\tau_{IIB}+b\over c\tau_{IIB}+d}, a,b,c,d \in {\bf Z},
\ee
and
\be
{B_{\mu\nu}\choose \tilde B_{\mu\nu}}= \left(\matrix{
a&b\cr
c&d\cr}\right){B_{\mu\nu}\choose \tilde B_{\mu\nu}},
\ee
 where the integers a,b,c,d are such that $ ab-cd=1$.\\
Following Vafa, one may interpret the complex field $\tau_{IIB}$ as the complex structure of an extra torus $T^2$ which, combined with the ten space time dimensions, leads to a twelve dimensional theory. From this view, $10d$ type IIB string may be seen as the compactification of F-theory on the elliptic curve $T^2$. Starting from F-theory one can do better by looking for new string models in lower dimensions obtained by compactifications on elliptically fibered Calabi-Yau manifolds. For example the eight dimensional F-theory on elliptically fibered $K3$ is obtained by taking the four compact variables as;
\be
 y^2=x^3+f(z)x+g(z),
\ee
 where $f$ and $g$ are polynomials of degree 8 and 12 in $z$ respectively.
  Eq (4.19) corresponds to vary the $\tau$ torus over the points $z$  of a compact space, which is taken to be a Riemann sphere $P^1$ parametrized by the local coordinates $z$. In other words the two-torus complex structure $\tau (z)$ is now a function of $z$ as it varies over the $P^1$ base of $K3$. Note alos  that eq(4.9) has generically 24 singular points corresponding to  $\tau (z)\to \infty$. These singularities have a remarkable physical interpretation. To each one of the 24 points, it is associated the location of a 7-brane of non perturbative type IIB string theory. Note moreover that F-theory on $K3$ is conjectured to be dual the heterotic string theory on $T^2$, with the heterotic coupling constant $g_s^h$ is given by the size of the $P^1$. This eight dimensional vacuum can be further compactified to lower dimensions, in particular to six dimensions. If we consider the compactification on an extra $T^2$, then the resulting duality becomes a duality between F-theory on K3$\t T^2$ and heterotic string on $T^4$ . The later is also known to be dual to type IIA string on K3 \cite{ht,wd,asp}. Like for the type IIA string, F-theory on local Calabi-Yau manifolds with ADE singularities give rise to non abelian ADE gauge groups. In what follows, we review the $N=1 $ $8d $ ADE  models from F-theory compactification on elliptically fibered $K3$;  then we study the extensions of the ADE construction to all non simply laced BCFG gauge groups, including affine and ordinary symmetries.
\subsubsection{$N=1$ 8d ADE models}
 In eight dimensions, F theory vacuum is determined by the moduli space of elliptically fibered K3 surface, while the gauge group is determined as usual by the affine ADE singularities of K3. Extending the geometric engineering approach of $4d$ $N=2$ superconformal QFT, the toric graphs of the blow up of the ADE singularities of the local elliptically fibered K3 surfaces are represented by a polyhedron $\Delta$(ADE) spanned by $(k=r+5)$ vertices of the standard lattice $Z^4$. These $(r+5)$ vertices $v_i, i=0,\ldots,r+4$ fulfill $( r+1)$ relations given by:
 \begin{equation}
\sum\limits_{i=0}^{r+4} q^a_iv_i=0,\quad  a=0,\ldots,r
\end{equation}
where $r$ is the rank of ADE algebras in question and  the $q^a_i$'s are the entries of the Mori vectors $q^a$  satisfying  the Calabi-Yau condition $\sum\limits_{i=0}^{r+4} q^a_i=0 $. The $q^a_i$ ' s give the intersection matrix of the $P^1$ curves of ADE toric polytopes and constitute with the $v_i$ vertices the toric data of the resolution of singularity. In the mirror space $W_2$, the resolution of the singularity is given by a dual polytope $TG_n$ represented by $k^*$ vertices $v_j^*$ satisfying similar relations as eq(4.7) with Mori vectors ${q^*}^a_j$. Following \cite{kmv,bm}, the manifolds $M_2 $ and $W_2$ are respectively given by the zeros of the following polynomials:
\begin{equation}
\begin{array}{lcr}
M_2=p_\Delta=\sum \limits_j a_j \prod_i x_i^{(v_i,v_j^*)+1}\\

W_2=p_\Delta^*=\sum\limits_ib_i\prod_jx_j^{(v_i,v_j^*)+1},
\end{array}
\end{equation}
where the sum (product) runs over all vertices $v_i$ $( v^* _j)$ and $a_j$  $(b_i)$ determine the complex structure of
$M_2$ $(W_2)$. In the present paper, we will focus our attention on the mirror geometry. Using the convention notation $ {(V_i)_j=(v_i,v^*_j)}$ which, up to a gauge fixing, can be viewed as the entries of $( r+1)$ three dimensional vectors $V_i=(s_i,n_i,m_i) $, where the first entry takes either zero or the value of the Dynkin weight of the adjoint representation of ADE Lie algebra. In other words the $r+5 $  vertices $ V_i$ belong to different $Z^2$ square sublattices of $Z^3$  depending on the values of $ s_i$. In the case we will be intersted in here, the $(r+5)$ vertices split in two cathegories :\\
(a)- Four vertices denoted as $\tilde V_l$ ; $ l= r+1, \ldots,r+4$ and whose entries are as:
\begin{equation}
V_l=(0,n_l,m_l).
\end{equation}
They define the elliptic curve E of the elliptic fibration in which the singularities are blown up.\\
(b)- $ r+1$  vertices
\begin{equation}
V_a=(s_a,n_a,m_a), \quad a= 0,1,  \ldots, r,
\end{equation}
representing the blown up curves of the resolution of the singularity. Obviously the $ V_i$ vertices of the polytope $ \Delta$ of the smooth space are not arbitrary; their entries satisfy:
\begin{equation}
\sum\limits_{i=0}^r q^a_iV_i +\sum\limits_{i=r+1}^{r+4} q^a_iV_i=0.
\end{equation}
Observe that this eq may be split as:
\begin{equation}
\begin{array}{clr}
 \sum \limits _{i=0}^{r+4} q^a_i s_i&=&0\\
\sum \limits _{i=0}^{r+4} q^a_i n_i&=&0\\
\sum \limits _{i=0}^{r+4} q^a_i m_i&=&0. \\
\end {array}
\end{equation}
To work out the explicit form of the the complex 2 dimension surface
 \begin{equation}
p(W_2) =\sum \limits  _{i=0}^{r+4} a_iy_i=0,
\end{equation}
defining the mirror geometry of the blown up singularity, we have to solve the following identity
\begin{equation}
\prod\limits_{i=0}^{r+4}y_i^{q^a_i}=1\qquad  a=0,\ldots,r,
\end{equation}
in terms of the local coordinates of the elliptic curve E.  To solve these eqs, we first associate to a vertex $V_i $ the gauge invariant monomial $y_i$;
 \begin {equation}
y_i={U_1}^{n_i}{U_2}^{m_i}{U_3}^{s_i};
\end{equation}
where $U_1$, $U_2$ and $U_3$ are three gauge invariant variables under the $C^*$ - action of the ambient weighted projective space $WP^3$ in which E is embedded. They will be specified later on. Note that eq(4.14) is trivially solved by the parameterization (4.15) so that eqs (4.13) may be replaced by,
\begin {equation}
W_2=  \sum\limits _{i=0}^{r+4} a_i{U_1}^{n_i}{U_2}^{m_i}{U_3}^{s_i},
\end {equation}
or equivalently,
\begin {equation}
W_2= \sum\limits_{l=r+1}^{r+4}a_l {U_1}^{n_l}{U_2}^{m_l}+ \sum\limits _{a=0}^{r}b_a{U_1}^{n_a}{U_2}^{m_a}{U_3}^{s_a}
\end {equation}
The second step in solving eqs(12) is to choose a polytope $\Delta$  for the elliptic curve E; i.e fix the values of the integers $n_l$ and $n_l$. In general, the choice of $n_l$  and $n_l$  is dictated by the type of the ADE singularity one considers. Nevertheless, to fix the ideas we make here the same choice as in \cite{kmv} and \cite{bfs2} by taking the elliptic curve E in the weighed projective space $ CP^2_ {1,2,3}$. Thus, the vertices
\begin{equation}
 \tilde V_{r+1}=(0,0,0), \tilde V_{r+2}=(0,-1,0) ,\tilde V_{r+3}= (0,0,-1) ,\tilde V_{r+4}=(0,2,3).
\end {equation}
 and the two gauge invariant variables under the projective $ C^*$ action of the space $ CP^2_ {1,2,3}$ are as follows \cite{bfs2}:
   \begin{equation}
 U_1={{zy}\over {x^2}}, \quad U_2={{xz}\over {y}}
\end{equation}
The remaining variable $U_3$ may be also introduced in the same manner as $U_1$ and $U_2$ by embedding $WP^2(1,2,3)$  in  $WP^3(1,2,3,\eta)$ of homogenous coordinates  $(z,x,y,w)$. Setting
 \begin{equation}
U_3= w f(z,x,y),
\end{equation}
where $ f(z,x,y)$ is a homogenous function of degre  $-\eta$, then putting eqs (4.19) and (4.20) in eq (4.17) and multiplying by $xyz$, one obtains the structure of two dimensional geometries which are the mirror of the local ADE singularites in the elliptic fibration over the complex plane.  The mirror geometry is given as a hypersurface defined by homogenous polynomial in some weighted projective spaces $WP^3$. Its general form is,
\begin{equation}
W_2= (y^2+x^3+z^6+ \mu  xyz)+ \sum\limits  _{i=0}^r a_i (wf)^{s_i}x^{1-2n_i+m_i}y^{1+n_i-m_i}z^{1+n_i+m_i}=0,
\end {equation}
 where the $w$ independent term describes the eplliptic curve with complex structure modulus $ \mu$, the $ w$ dependent term involves the ADE group and the $a_i$'s are complex parameters carrying the complex structure of $ W_2$.\\
For affine $ \hat A_n $, $ \hat D_n $ and $ \hat E_8 $ resolved singularities where it is convenient to use the elliptic curve E as in eq (4.18), the homogenous functions read as:
\begin{equation}
\begin{array}{clr}
 \hat A_n:\quad f(x,y,z)= z^{n-6}\\
\hat D_n:\quad  f(x,y,z)= z^{n-3}\\
\hat E_8:\quad f(x,y,z)= z^{-1}.
\end{array}
\end {equation}
 In these cases, the weighted projective spaces are $ WP_{1,2,3,6-n}$ , $ WP_{1,2,3,3-n}$  and  $ WP_{1,2,3,1}$  for $ \hat A_n $, $ \hat D_n $ and $ \hat E_8 $ respectively.

\section {N=1 8d BCFG models}
In  this section we want to derive the algebraic equations of the surfaces $ W_2$ describing the mirror geometry of the blow ups of non simply laced BCFG affine singularities of elliptically fibered $K3$. These eqs
complete the results described previously and allow to recover
the solutions for the moduli space of the Coulomb branch of $4d$ $N= 2$
superconformal gauge theories including non simply laced geometries \cite{bfs2}. The method we will be using to approach non simply laced symmetries is based on the standard
techniques of folding the Dynkin nodes of ADE graphs which are permuted by
the outer-automorphism groups $\Gamma$. Thus, starting form the ADE simply laced
geometries considered earlier, then folding the nodes of the toric graphs $TG$ that are permuted by $\Gamma$, one gets the constraints of non simply laced geometries. The general result is as follows:
\begin{equation}
\begin{array}{lcr}
 D_{n+1}/Z_2&\to B_n\\
A_{2n-1}/Z_2&\to C_n\\
E_6/Z_2&\to F_4\\
D_4/Z_2 &\to G_2.
\end{array}
\end {equation}
\subsection{Outer automorphism group of the toric diagram}
Ordinary and affine Lie algebras are
classified in two types: simply laced ADE algebras having a symmetric Cartan
matrix $K_{ij}=\alpha_i\alpha_j$ and non simply laced BCFG ones having a non
symmetric Cartan matrix $K_{ij}=2{\alpha_i\alpha_j\over\alpha_j^2 }$. This feature
is the main reason behind the complexity of the analysis of the blow ups of
the BCFG singularities. We see in this subsection how to deal with
this complexity; in particular how to extend the standard toric analysis to
non simply laced geometries by solving the constraints obtained by using folding techniques. In this approach, non simply laced Dynkin
diagrams $DG_{ns}$ may be obtained from the simply laced ones $DG_s$ by
identifying the Dynkin nodes which are permuted by the outer-automorphism
group $\Gamma$. Formally we may write this correspondence as:
\begin{%
equation} DG_{ns}\cong DG_{ns}/\Gamma .
\end{equation}
In the geometric
engineering of $4d$ $ N=2$ supersymmetric QFT in four dimensions with non simply laced gauge groups where the Dynkin
diagram $DG$ appears as a part of the toric graph $TG$ of mirror geometry, one should worry about the action of $\Gamma$ on the complementary part the
simply laced toric graphs $TG_s$. Extending eq(5.2) to the case of non
simply laced mirror diagram $TG_{ns}$, that is,
 \begin{equation}
TG_{ns}\cong
TG_{s}/ \Gamma,
\end{equation}
where $ TG_s=DG_s\cup G_0$, with $ G_0$ the
complementary part of $TG_s$ to be precised later on, one should distinguish
the two following cases :
\\ 1- $\Gamma$ acts trivially on $ G_0$ that is $
G_0/ \Gamma \cong G_0 $ and consequently:
\begin{equation} TG_{ns}\cong
DG_{ns}\cup G_0.
\end{equation}
 2-$\Gamma$ acts non trivially on $ G_0$ i.e $
G_0/ \Gamma \not\cong G_0 $ and so:
 \begin{equation} TG_{ns}\cong
DG_{ns}\cup (G_0/\Gamma).
\end{equation}
Moreover, once the action of on $G_0$
is specified; that is, $G_0$ stable or unstable under $\Gamma$, one has to
solve the action on the toric data of ADE the geometries. We shall turn to this question in moment; for the time being let
us discuss hereafter the part $DG_s$ ( $DG_{ns}$ ) of the
simply laced toric graphs $TG_s$ (non simply laced ones $TG_{ns}$). In what follows, we limit our presentation to the simply laced Dynkin diagrams having non trivial outer-automorphism groups. This leads us to discuss some topological features of the blow ups by complex curves.
 \subsection{Degeneracy of complex curves }
 In the case of simply laced toric graphs $TG_s$, the blow
up curves of the ADE singularity is given by a set of intersecting
$P^1$ curves according to the topology of the ADE Dynkin diagrams. Each $P^1$ curve intersects the adjacent ones at a point $(x, y, z)$ of $WP^2$ in which the $P^1$'s are embedded. From the topological point of view, $TG_s$
may be regarded as a particular limit of a connected sum of 2-spheres that
we describe herebelow; figures (1-6). Recall that a connected sum of
2-spheres, which topologically is also a 2-sphere, is given by gluing a disc
$D_1$ of the first sphere with an other disc $D_2$ of the second one.   $D_1$ and $D_2$
constitute then one disc $D$ only as shown on figure(1).
\begin{figure} [htpb]
\begin{center}
\begin {tabular}{c}
  \includegraphics[width=4cm , height =4cm]{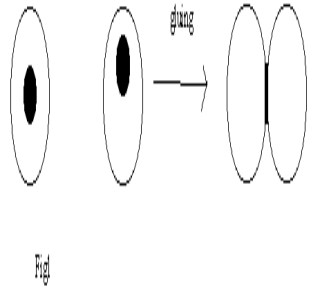}

\end{tabular}
\end{center}
\caption{\small  The gluing  of  two  spheres  by  Disk }
\end{figure}

 In the limit where $D$
reduces to a point $\{p=(x,y,z)\}$, one obtains then the topology of the blow up
2-spheres appearing in $ TG_s$ . We shall refer hereafter to the limiting case where
the disk $D$ of the connected sum reduces to a set of isolated points $\{p_i=(x_i,y_i,z_i)\}$ $D$ as: " {\it singular connected sum}". In figure(2), we represent two singular limits of the disc giving two special manifolds:\\
\begin{figure} [htpb]
\begin{center}
\begin {tabular}{c}
\includegraphics[width=4cm , height =4cm]{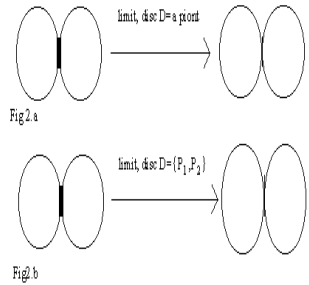}
\end{tabular}
\end{center}
\caption{\small  The two  special cases of  gluing limit of  two
spheres by Disk }
\end{figure}

{\it 1- the disc $D$ reduces to a point}\\
 In this case the resulting geometry is just the standard simply laced geometries of ADE geometries, see fig(2.a).\\
{\it {2- $D$ reduces to two isolated points $\{p_1=(x_1,y_1,z_1)\}$ and $
\{p_2=(x_2,y_2,z_2)\}$ of $ WP^2$}} \\
 The obtained geometry is as in fig(2.b).
In the language of toric geometry, the degeneracies of the connected sum
of the two 2-spheres have the mirror geometries represented by figure(3).
\begin{figure} [htpb]
\begin{center}
\begin {tabular}{c}
  \includegraphics[width=4cm , height =4cm]{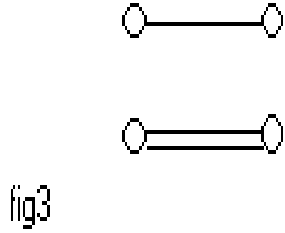}

\end{tabular}
\end{center}
\caption{\small The toric geometry of  the connected sum of two spheres }
\end{figure}

The nodes of the
graphs represent the compact 2-spheres of figures(2.a-b) and the links represent the intersecting points; a link for each point.
 \subsection{Non simply laced toric graphs}
First of all observe that the main difference between simply
laced toric graphs $TG_s$ and non simply laced ones $TG_{ns}$ is that in the
second geometry some of the blow up complex curves do not split completely. A collection of such real two-spheres which are not split corresponds effectively
to an orbit of $\Gamma$ obtained from a simply laced geometry by
imposing foldings of the 2-spheres permuted under $\Gamma$. Topologically, the
orbit of the permuted 2-spheres under $\Gamma$ means that the one complex
dimensional space one gets is not a single curve $P_1$; it is a union of $
P_1$'s with more than one intersecting point as shown on figure(2-b). In other
words, the toric geometry of the mirror blow up spheres of the orbit is
then as shown on figure(4).
\begin{figure} [htpb]
\begin{center}
\begin {tabular}{c}
  \includegraphics[width=4cm , height =4cm]{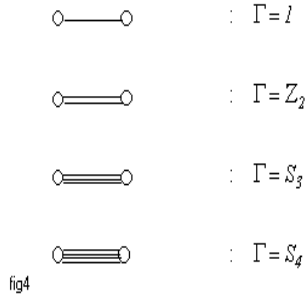}

\end{tabular}
\end{center}
\caption{\small The toric geometry of the mirror blow-up curves }
\end{figure}

From figure(5), one sees moreover that for a trivial group $\Gamma$, any
two blow up curves have one intersecting point; while for $\Gamma=Z_2$, $
\Gamma=S_3$ and $\Gamma=S_4$ one has respectively two, three and four intersecting points. Geometrically this means that
the Calabi Yau threefold has either no branch cut for $\Gamma=id$, one branch cut
for $\Gamma=Z_2$, two branch cuts for $\Gamma=S_3$ and three branch cuts
for $\Gamma=S_4$.
\begin{figure} [htpb]
\begin{center}
\begin {tabular}{c}
  \includegraphics[width=4cm , height =4cm]{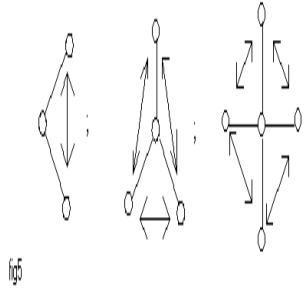}

\end{tabular}
\end{center}
\caption{\small The folding representations of spheres permuted  by  $\Gamma$ }
\end{figure}

More details will be given later on.
\subsection {Non simply laced BCFG affine mirror surfaces}
To describe the non simply laced mirror geometries of the ALE space with affine BCFG singularities, we proceed as follows:\\
 (i) solve the folding contraint equations by using the toric data of affine ADE singularuities.\\
(ii)Apply local mirror symmetry.\\
 The mirror geometry we obtain and which we denote as $ W_2$, is roughly speaking a local elliptically fibered  $K3$ surface with branch cuts.
 To derive the mirror potential associated with $ W_2$, we start by considering the contraint equation(5.4) and fix our attention on the two trivalent vertices of the affine $ D_n$ geometry figure(6).
\begin{figure} [htpb]
\begin{center}
\begin {tabular}{c}
  \includegraphics[width=4cm , height =4cm]{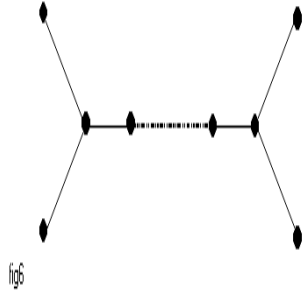}

\end{tabular}
\end{center}
\caption{\small The  affine  $ \hat D_n$ Dynkinn diagramm }
\end{figure}

 This geometry is specified by giving the two following data: First, the toric data $\{q^a_i,V_i\}$ of affine $D_n$ satisfying:
 \begin{equation}
 \begin {array}{lcr} \sum\limits _{i=0}^r q^a_i
V_i+\sum\limits _{l=r+1}^{r+4} q^a_l V_l=0 \\ \sum\limits _{i=0}^r
q^a_i+\sum\limits _{l=r+1}^{r+4} q^a_l=0.
\end{array}
\end{equation}
 Second, data concerning the actions of the outer-automorphism group $\Gamma=Z_2$ of the Dynkin diagram.\\
The topology of the $ \hat D_n$ graph shows that there are different possible realizations of $\Gamma$ leading to, amongst others, to the following two \cite{mdiagram}:(1)untwisted $B^{(1)}_{n-1}$.(2)twisted $B^{(2)}_{n-2}$. Without loss of generality we shall
discuss hereafter the first case in order to illustrate the idea.\\

  {\it{ $\hat  D_{n}$  foldes to  $\hat B^{(1)}_{n-1}$}}

 Up on folding the vertices $V_n$ and $V_{n-1}$ ;i.e,
\begin{equation}
 \begin{array}{lcr}
 V_n \leftrightarrow V_{n-1}\\ V_a \leftrightarrow V_a, a=0, \ldots,n-2,
\end{array}
 \end{equation}
it is not difficult to see that, if one forgets about non compact cycles, the resulting diagram is just the affine $\hat B_{n-1}$ Dynkin one. To build mirror geometry, this is however not the full story as we have to specify also the action of auter-automrphisms on the vertices of the elliptic curve E. Until now, we do not know how $\Gamma$ acts on the four vertices $\tilde V_l$ associated with E. All we is that the $\tilde V_l$'s are parts of the affine $\hat D_n$ mirror geometry. This is why we shall
extend the action of $\Gamma$ to the full toric graph and distinguish two
possible representations depending on whether the vertices of the
curve E are stable or unstable under $\Gamma$ . These two realizations of $
\Gamma$ are formulated as follows:\\
(a) The vertices of the curve E are
invariant under $ \Gamma$:\\
\begin{equation}
\begin {array}{lcr}
V_n \leftrightarrow V_{n-1}\\ V_a \leftrightarrow V_a,\quad  a=0, \ldots,n-2\\ \tilde V_l
\leftrightarrow \tilde V_l, \quad l=r+1, \ldots,r+4
\end{array}
\end{equation}
(b) The vertices of E are unstable under $\Gamma$. \\
In addition to the doublet $
(V_n,V_{n-1})$, the vertices $ \tilde V_{n+1}$ and $ \tilde V_{n+2}$ ($
\tilde V_{n+3}$ ) for $n$ even (odd) are also permuted under the action of $ \Gamma$ on the affine $\hat D_n$ toric graph:
\begin{equation}
 \begin{array}{lcr}
 V_n \leftrightarrow V_{n-1}\\ V_a \leftrightarrow V_a, \quad a=0,
\ldots,n-2\\ \tilde V_l \leftrightarrow \tilde V_l,\quad  l=n+3( n+2) , ,n+4\\ \tilde
V_{n+1} \rightarrow \tilde V_{n+2}(\tilde V_{n+3}).
 \end{array}
 \end{equation}
 In what follows we consider the representation (5.4) of $\Gamma$ in
order to state the constraint eqs(5.8) one obtains by folding the doublet $
(V_n,V_{n-1})$ of the affine $\hat D_n$ geometry. The other representation will be discussed in subsection 5.6
\subsection{ First class of solutions }
Under folding, the doublet $(V_n,V_{n-1})$
becomes a scalar $ V'_n$ and the old data $\{ q_i^a, V_i\}$ of the permuted
blow up spheres should be now replaced by new data $\{ {q'}_i^a, {V'}_i\}$ such that
eq (5.6) is still fulfilled. Moreover, the new Mori vector entries $q_j^a$ should be the
same as $ q^a_j$ for the values $a=0\ldots, n-3$ since the vertices $V_a, a=0,\ldots,n-3$ are invariant under $\Gamma$. Only the Mori vectors $
q^a, a=n-2,n-1,n$ which are affected. The folding constraint eqs are
then:
 \begin{equation}
 \begin {array}{lcr}
\sum\limits _{j=0}^{n+3} {q^c_j}
V'_j=0 \\ \sum\limits _{j=0}^{n+3} q^c_j=0,\quad  c=0,\ldots,n-1,
 \end{array}
\end{equation}\
 or equivalently by using the affine $\hat D_n$ toric data as well as eq
(5.6) and the property $ {q'} _j^a=q _j^a$ for $ a=0,1,\ldots,n-3$. We can then
rewrite the data (5.10) as follows:
 \begin{equation}
\begin {array}{lcr}
\sum\limits _{j=0}^{n+3} {q'}^{n-2}_j {V'}_j= \sum\limits _{j=0}^{n+3} {q'}^{n-2}_j
{V'}_j=0 \\ \sum\limits _{j=0}^{n+3} {q'}^{n-1}_j{V'}_j= \sum\limits _{j=0}^{n+4}
(q^{n-1}_j + q^{n}_j) V_j=0 \\ \sum\limits _{j=0}^{n+3} {q'}^a_j=0,
\quad a=0,1,\ldots,n-3 \\ \sum\limits _{j=0}^{n+3} {q'}^a_j=\sum\limits _{j=0}^{n+3}
q^a_j.,\quad a=0,\ldots, n-3.
\end{array}
 \end{equation}
 To determine the new data ${q'} ^a_j,\quad  a= n-2,n-1,n$
and ${ V'}_{n-1}$ as a function of the old one $\{q_j^a, V_j\}$, we should
solve the constraint eqs(5.11). Taking into account eq(5.6) and the explicit
expressions of the Mori vectors entries $ q_j^a$, eq(5.11) leads
to:
\begin{equation}
 \begin {array}{lcr}
 {q'}_i^{n-1}=q_i^{n-1}+q_i^n\\
{q'}_{n-1}^{n-1} {V'}_{n-1}=(q_{n-1}^{n-1} V_{n-1}+q_n^n V_n)+(q_{n-1}^n
V_{n-1}+q_n^{n-1} V_n)\\ {q'}_{n-1}^{n-2} {V}'_{n-1}=q_{n-1}^{n-2} V_{n-1}+q_n^{n
-2}V_n.
\end{array}
 \end{equation}
Moreover, using the fact that for affine $\hat D_n$
\begin{equation}
 \begin {array}{lcr}
q_{n-1}^{n-1}=q_n^n=-2,\quad q_{n-1}^{n-2}=q_n^{n-2}=1 \\ q_{n-2}^{n-1}
=q_{n-2}^{n}=q_{n+3}^{n-1}=q_{n+1}^{n}=q^n_{n+2}=1\\
q_{n}^{n-1}=q_{n-1}^n=q_{n+1}^{n-1}=q_{n+1}^{n}=q_{n+2}^{n-1}=0. \end{array}
 \end{equation}
 as well as (5.11), we find the following results:
\begin{equation}
\begin {array}{lcr}
 {q'}_{n-1}^{n-2}{V'}_{n-1}=V_{n-1}+V_n\\
{q'}_{n-1}^{n-1}{V'}_{n-1}=-2 (V_{n-1}+V_n)
 \end{array}
\end{equation}
and
 \begin{equation}
 \begin {array}{lcr}
 {q'}_j^{n-2}=-2\delta_j^{n-2}+\delta_{j+1}^{n-2}+2
\delta_{j-1}^{n-2}-\delta_{j-2}^{n-2}\\
{q'}_j^{n-2}=-2\delta_j^{n-1}+\delta_{j+1}^{n-1}-\delta_{j-1}^{n-1}+
\delta_{j-2}^{n-1}+\delta_{j-3}^{n-1}.
\end{array}
\end{equation}
Using these equations we find the following solution:
 \begin{equation}
 \begin {array}{lcr}
{ V'}_a=V_a,\quad  a=0,\ldots,n-2\\{ V'}_{n-1}={1\over
2}(V_{n-1}+V_n)\\ \tilde V_l= \tilde V_l,\quad l=n,\ldots,n+3\\
{q'}_j^a=q_j^a,\quad a=0,\ldots,n-3.
 \end{array}
 \end{equation}
Having determined the toric data of the affine $B^{(1)}_{n-1}$ blown up singularity, we can derive
the equation of the hypersurface $W_2$ of the affine $B^{(1)}_{n-1}$ mirror
geometry by using eq(5.16). Our results are:\\
(a)Toric vertices\\
\begin{equation}
\begin {array}{lcr}
V_0=(1,2,3),
V_1=(1,1,1)\\V_i=(2,3-i,4-i),\quad i=1,\ldots,n-3\\V_{n-1}=(1,{(5-n)\over 2
},{(6-n)\over 2 }) \\ \tilde V_n=(0,0,0),\tilde V_{n+1}=(0,-1,0),\tilde
V_{n+2}=(0,0,-1),\tilde V_{n+3}=(0,2,3).
 \end{array}
 \end{equation}
(b)The geometry $W_2$\\
 \begin{equation}
 \begin {array}{lcr}
P(W_2)=
(1+U_1+U_2+U_1^2U_2^3)\\+ U_3( a_0U_1^2U_2^3+ a_1 U_1U_2+a_{n-1}U_1
^{{(5-n)\over 2 }}U_2^{{(6-n)\over 2 }})+ U_3^2\sum\limits _{i=0}^{n-4}
a_{i+1}U_1^{3-i}U_2^{4-i}.
 \end{array}
\end{equation}
Replacing $U_1$,$ U_2$
and $U_3$ by their expressions eqs (4.19-20) and multiplying by $xyz$, one gets the
following complex two dimension surface of the the affine $ B_{n-1}^{(1)}
$ geometry:
 \begin{equation}
 \begin {array}{lcr}
 P(W_2)=
(y^2+x^3+z^6+ \mu xyz)+\\w(a_0 z^{n+3}+a_1 yz^n+a_{n-1} z ^{{7 \over2}}
x^{{(n-6)\over 2 }} y^{{1\over 2 }} ) + w ^2\sum\limits _{i=0}^{n-4} a_{i+1}
z^{2(n-i)} x^i.
 \end{array}
\end{equation}
 In what follows, we briefly
summarize our results for affine CFG non simply laced geometries associated with the first representation of the outer-automorphism groups $\Gamma$. We give
the toric data and the $P(W_2)$ eq of the surface for the untwisted
and twisted affine CFG resolved singularities.\\
{\bf (1) Untwisted $ C^{(1)}_n$ }\\
This geometry may be obtained from the $ SU(2n+1)$. The result is:\\
{\it (a) toric data}\\
 \begin{equation}
\begin {array}{lcr}
 V_i=(1,2-i,3-i), i= 0,n
;V_j=(1,2-j,{(5-2j)\over 2 }),j=1,\ldots,n-1 \\\tilde V_{n+1}=(0,0,0),\tilde V_{n+2}=(0,-1,0),\tilde
V_{n+3}=(0,0,-1),\tilde V_{n+4}=(0,2,3)\\
q^0_i=(-2,2,0^{n-1},-1,0,0,1),q^1_i=(1,-2,1,0^{n-2},-{1\over 2},0,{1\over
2},0)\\ q^2_i=(0^2,1,-2,1,0^{n-4},0,0,0,0), \ldots ,q^{n-2}_i=( 0^{n-3},1,
-2,1,0,0,0,0,0)\\ q^{n-1}_i=(0^{n-2},1,-2,1,-{1\over 2}, 0,{1\over 2},0),
q^n_i=(0^{n-1},2,-2,-3,2,1,0), \i=0,\ldots,n+4.
 \end{array}
 \end{equation}
 {\it$ (b) C^{(1)}_n$ mirror geometry}:\\
\begin{equation}
 \begin {array}{lcr}
P(W_2)= (y^2+x^3+z^6+ \mu xyz)+\\ + w (
a_0z^{2n}+\sum\limits _{i=1}^{n-1} a_{i} z^{{(4(n-i)-1)\over 2}} y^{{1\over
2}}x^{{( 2i-1)\over 2 }}) +a_nx^n.
\end{array}
 \end{equation}
{\bf (2) Untwisted $F^{(1)}_4$} \\
{\it (a) Toric data}\\
 \begin{equation}
 \begin {array}{lcr}
V_0=(1,-1,1),
V_1=(2,-1,0), V_2=(3,-1,-1), V_3=(2,-{1 \over 2},-1), V_4=(1,0,-1),\\ \tilde
V_5=(0,0,0),\tilde V_6=(0,-2,-1),\tilde V_7=(0,-1,2),\tilde V_8=(0,-1,-1)\\
q^0_i=(-2,1,0^5,1,0),q^1_i=(1,-2,1,0^6,)\\ q^2_i=(0,1,-2,2,0,-1,0,0,0),
q^3_i=( 0,0, 1,-2,1,0,0,0,0)\\ q^4_i=(0,0,0,1,-2,0 ,{1 \over 2} ,0,{1 \over
2}),\quad  i=0, \ldots,7 .
\end{array}
 \end{equation}
{\it (b) The mirror geometry}\\
 \begin{equation}
 \begin {array}{lcr}
 P(W_2)= (y^3+x^3+z^3+ \mu xyz)+ w(
y^2+xz)+w^2(y+x^{{1 \over 2}} z^{{1 \over 2}})+w^3.
 \end{array}
 \end{equation}
{\bf (3) Untwisted $\hat G^{(1)}_2$}\\
{\it (a) toric data}\\
 \begin{equation}
 \begin {array}{lcr}
V_0=( 1,2,3),
V_1=(2,2,3), V_2=(1,{2\over3},1), \\ \tilde V_3=(0,0,0),\tilde
V_4=(0,-1,0),\tilde V_5=(0,0,-1),\tilde V_6=(0,2,3)\\
q^0_i=(-2,1,0^4,1),q^1_i=(1,-2,3,-2,0,0,0)\\ q^2_i=(0,1,-2,-{2\over
3},{2\over 3},1,0),\quad i=0, \ldots,6 .
\end{array}
 \end{equation}
{\it (b) The mirror geometry}\\
 \begin{equation}
\begin {array}{lcr}
P(W_2)= (y^2+x^3+z^6+ \mu
xyz)+ w(z^7+ y^{{2\over 3}}x ^{{2\over 3}}z^{{11\over 3}})+w^2z^8.
 \end{array}
 \end{equation}
{\bf {(4) The toric data of twisted $B^{(2)}_{n-1}$}}\\
{\it (a) Toric data}\\
 \begin{equation}
\begin
{array}{lcr} V_0=(1,{3\over2} ,2), V_i=(2,3-i,4-i),i=1,\ldots,n-2
V_{n-1}=(1,{(5-n)\over 2},{(6-n)\over 2},\\ \tilde V_{n}=(0,0,0),\tilde
V_{n-1}=(0,-,10),\tilde V_{n+2}=(0,0,-),\tilde V_{n+3}=(0,2,3)\\
q^0_i=(-2,1,0^{n-3},0,0,{1\over 2},{1\over 2}),\quad q^1_i=(2,-2,1,0^{n-4},-1,0,0,0)\\
q^2_i=(0,1,-2,1,0^{n-5},0,0,0,0),\ldots, q^{n-3}_i=( 0^{n-4},1,-2,2,-1,0,0,0)\\
q^{n-2}_i=(0^{n-3},1,-2,-1,1,1 ,0)), i=0, \ldots,n+3.
 \end{array}
 \end{equation}
{\it (b) The mirror geometry}\\
 \begin{equation}
\begin {array}{lcr}
P(W_2)=
(y^2+x^3+z^6+ \mu xyz)+\\w(a_0 y^{{1 \over2}}z^{{(2n+3) \over 2}}+a_{n-1} z ^{{7 \over2}} x^{{(n-2)\over 2 }} y^{{1\over 2 }} ) + w
^2\sum\limits _{i=0}^{n-3} a_{i+1} z^{2(n-i)} x^i.
 \end{array}
\end{equation}
{\bf (5) Twisted $F^{(2)}_4$}\\
{\it (a) toric data}\\
 \begin{equation}
 \begin {array}{lcr}
V_0=(2,-1,0), V_1=(4,-2,-1), V_2=(3,{-{3 \over 2}},-1), V_3=(2,-1,-1),
V_4=(1,{-{1\over 2}},-1),\\ \tilde V_5=(0,0,0),\tilde V_6=(0,2,-1),\tilde
V_7=(0,0,1),\tilde V_8=(0,-2,1)\\ q^0_i=(-2,1,0^4,{1\over 2},0,{1\over 2}),\quad q^1_i=(1,-2,1,0^6,)\\
q^2_i=(0,1,-2,1,0,0,0,0,0), \quad q^3_i=( 0,0,2,-2,1,-1,0,0,0)\\ q^4_i=(0,0,0,1,-2,0
,0 ,0,1 0),\quad  i=0, \ldots,7 .
\end{array}
 \end{equation}
{\it (b) The mirror geometry}\\
\begin{equation}
\begin {array}{lcr}
 P(W_2)= (y^2+x^3+z^6+ \mu xyz)+ w(
x ^{{3\over 2}}z^{{3\over 2}})+w^2(y+xz) +w^3 x ^{{1\over
2}}z^{{1\over 2}}+w^4.
 \end{array}
 \end{equation}
{\bf (6) Twisted $\hat G^{(2)}_2$}\\
{\it (a) toric data}\\
 \begin{equation}
 \begin {array}{lcr}
V_0=( 1,-{1\over3},-{1\over3}),
V_1=(2,-{2\over3},-{2\over3}), V_2=(3,-1,-1), \\ \tilde V_3=(0,0,0),\tilde
V_4=(0,2,-1),\tilde V_5=(0,-1,2),\tilde V_6=(0,-1,-1)\\
q^0_i=(-2,1,0^4,1),q^1_i=(1,-2,3,-2,0,0,0)\\ q^2_i=(0,1,-2,-{2\over
3},{2\over 3},1,0,0),\quad i=0, \ldots,6 .
\end{array}
 \end{equation}
{\it (b) The mirror geometry}\\
 \begin{equation}
\begin {array}{lcr}
P(W_2)= (y^3+x^3+z^3+ \mu
xyz)+ w( y^{{2\over 3}}x ^{{2\over 3}}z^{{2\over 3}})+w^2(y^{{1\over 3}}x
^{{1\over 3}}z^{{1\over 3}})+w^3.
 \end{array}
 \end{equation}
 \subsection{ Second class of solutions }
Here we study the second representation of the outer-automorphism group $\Gamma$ given by eq(5.5). In this realization, $\Gamma$ acts not only on the vertices of the Dynkin part of the toric diagram but also on the toric data of the elliptic curve E. It
turns out that this representation has the feature of giving mirror
geometries of complex dimension one, that is one dimension less than the
natural representation (figure 4). This dimensional reduction can be easily seen
from the toric data $ \{q_i^a,V_i\}$ before the folding and data $ \{{q'}_i^a
 ,{V'}_i\}$ after folding. Before folding, the $(n+4)$ vertices
satisfy in general $n$ relations so that the complex dimension $ d=
((n+4)-n)-2=2$. However after folding using the representation eqs(5.5) and
 figure 5, there are only $(n+2)$ vertices and $(n-1)$ relations; the complex dimension $d
$ which is given by $ d= ((n+2)-(n-1))-2$ is equal to one. This is however not a problem as one
may usually introduce extra auxliary variables to restore the correct dimensions without affecting the complex structure of the
mirror geometries. The constraint eqs of folding under the $ \Gamma$
representation figure 5 can be easily derived from the elliptic curve
E eq. However since under folding, two vertices $ \tilde V_i$ and $ \tilde V_j$ of
E are in the same orbit of $\Gamma$, we get extra conditions. In the case of the
affine  $\hat D_n$ with $n$ odd for example, where $V_{n+1}$ and $V_{n+3}$ transform as a doublet under $\Gamma$, the
constraint reads as:
\be
U_2=1
\ee
or equivalently
\be
y=xz
\ee
Putting back the constraint eqs
(5.32) and (5.33) into eqs (4.17) and (4.21), one obtains the equation of the mirror geometry we are looking for: \\
\be
P(W_2)= (x^3+z^6+ \mu
x^2z^2)+ w(a_0z^{n+3}+a_1xz^{n+1}+a_{n-1}z^4x^{{(n-1)\over 2}})+w^2\sum\limits _{i=0}^{n} a_{i+2} z^{2(n-i)} x^i.
\ee
In what follows we collect the non trivial results we have obtained.\\
{\bf (1) Untwisted $C^{(1)}_n$} \\
{\it (a) toric data}\\
\begin{equation}
\begin {array}{lcr}
V_i=(1,2-i),\quad  i=0,\ldots,n \\\tilde V_{n+1}=(0,0),\quad \tilde
V_{n+2}=(0,-1),\quad \tilde V_{n+3}=(0,2),\\
q^0_i=(-2,2,0^{n-1},-1,0,1),\quad q^1_i=(1,-2,1,0^{n+1})\\
q^{n-1}_i=(0^{n-2},1-2,1,0,0,0), \ldots ,q^{n}_i=( 0^{n-1},2,-2,-2,2,0)
\end{array}
\end{equation}
{\it (b) Mirror geometry} \\
\begin{equation}
\begin {array}{lcr}
P(W_2)= (x^3+z^6+ \mu
x^2z^2)+w\sum\limits _{i=0}^{n} a_{i} z^{2(n-i)} x^i.
\end{array}
 \end{%
equation}
{\bf (2) Untwisted $ F^{(1)}_4$} \\
{\it (a) toric data}\\
 \begin{equation}
 \begin {array}{lcr}
V_0=(1,1), V_1=(2,0), V_2=(3,-1), V_3=(2,-1),
V_4=(1,-1),\\ \tilde V_5=(0,,0),\tilde V_6=(0,-1),\tilde
V_7=(0,2),\\ q^0_i=(-2,1,0^5,1),q^1_i=(1,-2,1,0^5)\\
q^2_i=(0,1,-2,2,0,-1,0,0), q^3_i=( 0,0,1,-2,1,0,0,0)\\ q^4_i=(0,0,0,1,-2,0
,1 0),\quad  i=0, \ldots,7 .
\end{array}
 \end{equation}
{\it (b) Mirror geometry}\\
 \begin{equation}
\begin {array}{lcr}
 P(W_2)= (y^3+x^3+ x^2y)+ w(
x ^2+y^2)+w^2(y+x )+w^3,\quad a_i=1
 \end{array}
 \end{equation}
{\bf (5) Untwisted $\hat G^{(1)}_2$}\\
{\it (a) toric data}\\
 \begin{equation}
 \begin {array}{lcr}
V_0=( 1,1),
V_1=(2,0), V_2=(3,-1), \\ \tilde V_3=(0,1),\\
\end{array}
 \end{equation}
{\it (b) Mirror geometry}
 \begin{equation}
\begin {array}{lcr}
P(W_2)= x^3)+ a_0w^3+a_1w^2x+a_2wx^2.
 \end{array}
 \end{equation}
So far we have been studying  $N=1$ supersymmetric gauge theory in eight dimensions. However under further compactifications, we get in general extended supersymmetric quantum field theories in reduced dimensions. One of the interesting examples is the compactifications down to four dimensions. In what follows we want to extend the previous results to all ordinary non simply laced geometries using type IIA string compactification on elliptically fibered Calabi-Yau threefolds.\\
 \section {Ordinary geometries from type II strings}
\subsection {Toric data for type IIA ordinary ADE geometries}
 We begin by giving
a rapid review on toric data of ordinary ordinary ADE geometries used in the geometric engineering method of $ N=2 $ $4d$ supersymmetric QFT. In addition to bivalent vertices, these
geometries involve also the so called trivalent $ T_{p,q,r}$
geometry \cite{kmv}. As a basic result, it is well known that
$D_n$, $E_6$, $ E_7$ and $E_8$ type IIB mirror geometries correspond
respectively to $ T_{2,2,n-2}$ ,$ T_{2,3,3}$ , $ T_{2,3,4}$ and $ T_{2,3,5}$. We shall give hereafter the toric data for these geometries and for completeness, we give also the mirror of ordinary $ A_n$ using only  bivalent geometry.\\
 {\bf $ A_n$ geometry}: \\
The toric data of
this background is represented by $ n+2$ vertices $ V_i$ of the $ Z^2$ lattice fulfilling the following $ n$ relations:
 \begin{equation}
 \begin {array}{lcr}
\sum \limits _{i=0}^{ n+1}q^a_iV_i=0,
\end{array}
\end{equation}
 where,
\begin{%
equation} \begin {array}{lcr}
V_i=(1,i),\quad  i=0,\ldots, n+1 \\
q^a_i=(0^{i-1},1,-2,1,0^{n-i}),\quad  a=1,\ldots, n.
 \end{array}
 \end{equation}
The type IIB mirror description of this space is given, after solving the equations (2.14-15), by:
\begin{equation}
 \sum\limits _{i=0}^{n+1}a_iz^i=0,\qquad
a_0=a_{n+1}=1 .
\end{equation}
 Note that this mirror geometry is a $0$ dimensional space; this ambiguity may be rised by using the trick of auxiliary variables as we have done in different occasions throughout this paper as in eq(2.15) for example. \\
{\bf  $D$ and $E$ geometries}\\
The toric data of these background geometries are given in terms of
trivalent and bivalent vertices. \\
{\it (1)$ D_n$ toric data}\\
\begin{equation}
\begin {array}{lcr} \\
 V_i
=(0,0,n-2-i) ,V_{n-1}=(0,1,0), V_{n}=(1,1,0), \quad  i= 1,\ldots, n-2 \\\tilde
V_{n+1}=(1,1,1),\tilde V_{n+2}=(1,0,0),\tilde V_{n+3}=(0,1,0),\tilde
V_{n+4}=(0,0,1)\\ q^1_i=(-2,1,0^{n+1},-1),\quad q^a_i=(0^i,1,-2,1,0^{n+1-i}) ,
a=2, \ldots , n-3\\ q^{n-2}_i=(0^{n-4},1,-2,1,1,-1,0,0,0)\\
q^{n-1}_i=(0^{n-3},1,-2,0,0, 1,0,0), \\q^n_i=(0^{n-3},1,0,-2,0,0,1,0),\quad  i=1,
\ldots , n+4 .
\end{array}
\end{equation}
{\it (2)  toric data of $ E_{6,7,8}$} \\
\begin{%
equation} \begin {array}{lcr} E_6 :\\ V_1 =(0,0,2), V_2=(0,1,0), V_3=(0,0,1)
\\ V_4 =(0,0,0), V_5=(1,0,0), V_6=(2,0,0) \\ \tilde V_7=(1,1,1),\tilde
V_8=(3,0,0),\tilde V_9=(0,2,0),\tilde V_{n+4}=(0,0,3)\\ q^1_i=(-2,1,0,
0,0,0,0,0,0,1)\\ q^2_i=(1,-2,1, 0,0,0,0,0,0,0)\\ q^3_i=(0,1,-2,
1,0,0,0,0,0,0)\\ q^4_i=(-2,1,0, 0,0,0,0,0,0,0)\\ q^5_i=(0,0,0,
1,0,0,0,0,0,0)\\ q^6_i=(0,0,1, 0,0,-2,0,0,1,0)\\
\end {array}
\end{equation}
 \begin{equation}
\begin {array}{lcr} E_7:\\
 V_1 =(2,0,0),
V_2=(0,1,0), V_3=(1,0,0) \\ V_4 =(0,0,0), V_5=(1,0,0), V_6=(2,0,0), V_7=(3,0,0) \\ \tilde
V_8=(1,1,1),\tilde V_9=(4,0,0),\tilde V_{10}=(2,0,0),\tilde V_{11}=(3,0,0)\\
q^1_i=(-2,1,0, 0,0,0,0,0,0,0,1)\\ q^2_i=(1,-2,1, 0,0,0,0,0,0,0,0)\\
q^3_i=(0,1,-2, 1,0,0,1,-1,0,0,0)\\ q^4_i=(0,0,1,-2,1,0, 0,0,0,0,0)\\
q^5_i=(0,0,0, 1,-2,1,0,0,0,0,0)\\ q^6_i=(0,0,0,0,1,-2,0,0,1,0,0)\\
q^7_i=(0,0,1,0,0,0,-2,0,0,1,0).
\end {array}
\end {equation}
\begin{equation}
 \begin {array}{lcr} E_8 :\\ V_1 =(0,0,2),
V_2=(0,1,0), V_3=(0,0,1) \\ V_4 =(0,0,0), V_5=(1,0,0), V_6=(2,0,0),V_7=(3,0,0),V_8=(4,0,0) \\ \tilde
V_9=(1,1,1),\tilde V_8=(5,0,0),\tilde V_9=(0,2,0),\tilde V_{n+4}=(0,0,3)\\
q^1_i=(-2,1,0, 0,0,0,0,0,0,0,0,1)\\ q^2_i=(1,-2,1, 0,0,0,0,0,0,0,0,0)\\
q^3_i=(0,1,-2, 1,0,0,0,1,-1,0,0,0)\\ q^4_i=(0,0,1,-2,1,0, 0,0,0,0,0,0)\\
q^5_i=(0,0,0,1,-2,1,0,0,0,0,0)\\ q^6_i=(0,0,0,0,1, -2,1,0,0,0,0,0)\\
q^7_i=(0,0,0, 0,0,1,-2,0,0,1,0,0)\\ q^8_i=(0,0,1, 0,0,0,1,-2,0,0,1,0).
\end {array}
\end {equation}
 Having these toric data of the ordinary $D$ and $E_r$ backgrounds, one may determine the
mirror geometries by solving eqs(2.14-15). Our results are listed herebelow:
\begin{equation}
\begin{array}{lcr}
 D_n:\\ p(W_2)=( z^{n-2}+x^2+y^2+xyz )+\sum \limits
_{i=0}^{n-3}a_i z^i+a_{n-2}x+a_{n-1}y\\ E_6:\\ p(W_2)=( z^3+x^3+y^2+xyz )+(
a_0+ a_1 x+ a_2y+ az+ a_4x^2+ a_5z^2)\\ E_7:\\ p(W_2)=( z^3+x^4+y^2+xyz )+(
a_0+ a_1 x+ a_2y+ az+ a_4x^2+ a_5z^2+a_6 x^3) \\ E_8:\\ p(W_2)=(
z^3+x^5+y^2+xyz )+( a_0+ a_1 x+ a_2y+ az+ a_4x^2+ a_5z^2+a_6 x^3 +a_7 x^4)
\end{array} \end{equation}
 \subsection {Non simply laced ordinary mirror geometries}
Following the same method of folding we have used in the study of affine geometries one obtains in the context of F-theory compactification on elliptically fibered K3( section 5), we can determine the mirror  potentials for the non simply laced BCFG geometries for type II string compactifications. Our results are:\\
\subsection {First class of solutions for ordinary BCFG geometries}
 {\bf (1) $ B_{n-1}$case}\\
  {\it (a) toric data}\\
 \begin{equation} \begin {array}{lcr}
V_i=(0,0,n-2-i), i=1,\ldots, n-2, V_{n-1}=({1\over2},{1\over2}, 0)
\\ \tilde V_{n}=(1,1,1),\tilde V_{n+1}=(2,1,0),\tilde
V_{n+2}=(0,2,0),\tilde V_{n+3}=(0,0,n-2).
\end{array} \end{equation}
{\it (b) Mirror geometry}\\
 \begin{equation}
p(W_2)=( z^{n-2}+x^2+y^2+xyz )+ a_1
x^{1\over2}y^{1\over2}+\sum\limits _{i=1}^{n-2}a_{i+1}z^{{n-i-2\over 2}}.
\end{equation}
{\bf (2) $ F_4$}\\
{\it (a) toric data }\\
  \begin{equation}
\begin {array}{lcr} V_1=(0,0,1),
V_2=(0,0,0), V_3=({1\over2},{1\over2},0), V_4=(1,1,0),\\ \tilde
V_5=(1,1,1),\tilde V_6=(3,0,0),\tilde V_7=(0,3,0),\tilde V_8=(0,0,2)\\
q^0_i=(-2,1,0^5,1),q^1_i=(1,-2,1,0^5)\\ q^2_i=(0,1,-2,1,0,0,0,0), q^3_i=(
0,0,-2,1,-1,0,0,0)\\ q^4_i=(0,0,0,1,-2,0 ,0 ,1 ), i=0, \ldots,7.
\end{array} \end{equation}
 {\it (b) mirror geometry }\\
\begin{equation}
\begin {array}{lcr}
p(W_2)=( z^2+x^3+y^3+xyz )+ a_1z +a_2+a_3 xy+a_4x^{{1\over2}} y^{{1\over2}}
\end{array}
 \end{equation}
{\bf (3) Twisted $G^{(2)}_2$}\\
 {\it (a) toric data}\\
 \begin{equation}
\begin {array}{lcr}
V_1=(0,0,0), V_2=({1\over3},{1\over3},{1\over3}), \\
\tilde V_3=(1,1,1),\tilde V_4=(2,0,0),\tilde V_5=(0,2, ),\tilde
V_6=(0,0,2)\\ q^0_i=(-2,1,0^4,1),q^1_i=(1,-2,3,-2,0,0,0)\\
q^2_i=(0,1,-2,-{2\over 3},{2\over 3},1,0),i=0, \ldots,6
 \end{array}
\end{equation}
{\it (b) mirror geometry}\\
 \begin{equation}
 \begin {array}{lcr}
p(W_2)=(
z^2+x^2+y^2+xyz )+ a_1+a_2x^{{1\over3}} y^{{1\over3}} z^{{1\over3}}.\end{array%
} \end{equation}
\subsection {Second class of solutions for ordinary BCFG geometries}
{\bf (1) $ B_{n-1}$ case} \\
{\it (a) toric data}\\
 \begin{equation}
 \begin{array}{clr} V_i =
(1,n-2-i), V_{n-1}=(1,0),i=1,... ,n-2\\ V_n = (2,1), V_{n+1} = (2,0),
V_{n+2} = (0,n-2).
 \end{array} \end{equation}
 {\it (b) mirror geometry}\\
 \begin{equation}
P(W_2)= (z^{n-2}+x^2+x^2z) +\sum \limits _{i=0}^{n-3}a_i z^i +b_{n-1}x
\end{equation}
{\bf (2) $ C_n $ case }\\
{\it (a) toric data}\\
 \begin{equation} \begin{array}{clr}
V_i = (i,0),i=1,... ,n\\ \tilde V_{n+1} = (1,1), \tilde V_{n+2} = (0,0),
\tilde V_{n+3} = (3,1)
\end{array} \end{equation}
{\it (b) mirror geometry}\\
 \begin{equation}
P(W_2)= (1+xy+x^3y) +\sum \limits _{i=1}^{n}a_i x^i.
\end{equation}
{\bf (3) $ F_4$ case}\\
{\it (a) toric data}\\
\begin{equation}
\begin{array}{clr}
 V_1 =
(0,1),V_2 = (0,0),V_3 = (1,0),V_4 = (2,0)\\
\tilde V_{5} = (2,1), \tilde
V_{6} = (2,0), \tilde V_{7} = (3,0).
 \end{array}
\end{equation}
{\it (b) mirror geometry}\\
 \begin{equation}
P(W_2)= (z^2 +x^3+x^2z) +
a_0+a_1x+a_2x^2+a_3z.
\end{equation}
{\bf (4) $G_2$ case }\\
{\it (a) toric data}\\
 \begin{equation}
\begin{array}{clr}
V_1 = (1,1),V_2 = (0,1)\\ \tilde V_{3} = (3,1), \tilde
V_{4} = (2,1).
\end{array}
\end{equation}
{\it (b) mirror geometry}
\begin{equation}
 P(W_2)= (x^3y+x^2y) + a_0xy+a_1x.
 \end{equation}
In the end of this section, it interesting to note that bivalent and trivalent geometries describe all mirror complex two dimensions surfaces except affine $ \hat D_4$ which seems to need tetravalent geometry. Herebelow we want to study this special case.
 \section{Polyvalent geometry }
 In this section, we want to build mirror manifolds involving polyvalent toric geometry where the bivalent and trivalent ones appear just as the leading terms of a more general case. Roughly speaking polyvalent geometry may be viewed as an extention to higher dimensions of the bivalent and trivalent geometries we have considered earlier. Before discussing this extended geometry, it is useful to review some basic facts about the bivalent and trivalent ones.\\
\subsection{Bivalent geometry  }
Bivalent geometry, called also linear geometry, is based on bivalent vertices only and appears in the mirror of type IIA on $A_n$ space. In the geometric engineering of supersymmetric QFT, this geometry is used to describe a linear chain of gauge groups $ \prod SU $  with bi-fundamantal matters. It is also used in rederiving results of the Seiberg-Witten model from type IIA string on local $ K3$ fibered Calabi-Yau threefolds where the $K3$ surface has an $A_n$ singularity fibered on a complex $P^1$ curve . In toric geometry langauge, this means that the Mori vectors take the form:
\be
q^a_i=(-2,1,1,0,...,0),
\ee
 and the vertices are points of the $ {\bf Z}^2$ lattice.
 In few word, bivalent geometry may be thought of as a linear chain of  divisors with self intersection $(-2)$ and intersect the two adjacent divisors once with contribution $(+1)$. For more details see \cite{kmv,bfs1}
\subsection{Trivalent geometry}
 Trivalent geometry however involves both bivalent and trivalent vertices. The trivalent geometry has been used in diferent occasions. First it has been used in four dimensions to incorporate fundamental matters in a linear chain of $ \prod SU$ gauge group with $ N=2$ bifundamental matters \cite{kmv}. Second, trivalent geometry has also been used to describe the complex resolution of affine ADE singularities in a fibration of an elliptic curve over the complex plane of the local Calabi Yau threefolds. Note that these singularities play an improtant role in the geometric construction of  $ N=2$ superconformal theories in four dimensions. In toric geometry language, trivalent geomtry contains a central divisor with self intersection $(-2)$ intersecting three other divisors once with contribution $(+1)$. Put differently the Mori vector of the trivalent vertex is given by:
\be
q^a_i=(-2, 1, 1, 1,-1,0,...,0).
\ee
 The vertices $ v_i$ of the corresponding toric polytopes are given by points of the cubic lattice  ${\bf Z}^3$ satisfying eqs(2.8) as well as the Calabi-Yau condition $\sum q^a_i=0 $. The local mirror map of this geometry one has after solving eqs(2.14-15) contains the following monomials:
\be
1,x,y,z,xyz,
\ee
where $1$ corresponds to central divisor; $x$, $y$ and $z$ to divisors with contribution (+1 ) and $xyz$ to an extra vertex with contribution $(-1)$ in order to fulfill the Calabi-Yau condition: $ \sum q^a_i=0$. This geometry describes the complex deformation of the $T_{p,q,r}$ singularity defined as the intersection of three chains type $A_{p-1}$ $A_{q-1}$,$A_{r-1}$. More generally, the trivalent mirror geometry contains the monomials:
\be
1,x,y,z,xyz;x^2,x^3,....,x^p;y^2,y^3,...,y^q;z^2,z^3,...,z^r.
\ee
The latters are used in the blow up of elliptic $E_s; s=6,7,8$  singularities. A list of the mirror potentials of ADE geometries can be found in \cite{kmv,bfs2}.\\
\subsection{Extended  geometry}
Polyvalent geometry may be defined as based on a polyvalent divisor $D_0$ with self intersection $(-2)$ intersecting $n$ other divisors $ D_i,i=1,...,n$ with contribution $(+1)$. Like for the bi and trivalent case, this geometry is represented by a central vertex $ v_0$ connecting $n$ other vertices $ v_i$, which are represented by points in the hypercubic $
 Z^n$ lattice. Calabi-Yau condition requires that is the entries of each Mori vector $q^0_i$ associated with $ v_0$ should to add to zero. This implies in turns that one has to add an extra $D_{n+1}$ divisor with contribution $2-n$; i.e,
\be
q^0_i=(-2,1^n, 2-n),
\ee
or equivalently
\bea
D_0.D_0&=&-2\nn\\
D_0.D_i&=1&,\qquad i=1,\ldots,n\\
D_0.D_{n+1}&=&2-n\nn.
\eea
 The local mirror polyvalent geometry one gets after solving eqs(2.14-15), contains the following monomials;
\be
 1, x_1^{n-2},x_2^{n-2},... x_n^{n-2},\prod\limits_{i=1}^nx_i.
\ee
 For  $n=2$, we find the bivalent geometry and for $n=3$ we rediscover   the trivalent geometry.
 For $n=4 $ however, we get the missing tetravalent vertex of affine $\hat  D_4$ geometry. It is described by the following monomial.
\be
1,x_1^2,x_2^2,x_3^2,x_4^2,x_1x_2x_3x_4.
\ee
This geometry may be used to extend the $T_{p,q,r}$ singularity to $T_{p,q,r,t}$ by considering four intersecting $SU$ chains.\\

{\bf {Tetravalent geometry}}\\
 This is a particular example of polyvalent geometry which comes after the bivalent and trivalent geometries. Tetravalent geometry, which describes the  affine $ so(8)$  Dynkin diagram of the $N=2$ superconformal field theory in four dimensions, can be shown to have two toric realizations depending on whether we are using an integer or rational lattice. We shall refer hereafter to these two solutions as special tetravalent geometry and standard tetravalent geometry. \\

(1){\it { Special tetravalent geometry}}.\\
It is based on rational polyptopes $\Delta_{rational}$ of the cubic half integer $ {\bf Z}\times ({1\over2} {\bf Z}^2)$. We have found the following data for  $\Delta_{rational}$ \cite{bfs1}:
\bea
V_1=(1,1,1),\quad V_2=(1,2,3),\quad V_3=(1,{1\o 2},{1\o 2})\nn\\V_4=(1,{1\o 2},{3\o 2}),V_5=(2,2,3).
\eea
The mirror of the  blow up of affine $ \hat D_4$ singularity of the elliptic $ K3$ we have got is:
\be
W_2(x,y,z,w)=  (y^2+x^3+z^6+ \mu xyz)+ w(x^2z^4+z^8+x ^{{3\over 2}}z^5
+x ^{{1\over 2}}z^4y)+w^2z^{10}
\ee
$W_2(x,y,z,w)$ is a complex surface with double covers; this is price one should pay for embedding the mirror geometry in $WP^3$. This ambiguity is overpassed in standard geometry we want to introduce now.\\

(2) {\it {Standard tetravalent geometry}}.\\
Standard tetravalent geometry ia a natural extension of what we have in bivalent and trivalent cases. The correponding vertices are in the integer quartic $ Z^4$. The toric data one obtains are:
\bea
V_1=(1,1,-1,-1),
Vv_2=(1,-1,-1,1),
V_3=(2,-1,-1,-1),\nn\\
V_4=(1,-1,-1,-1),
V_5=(1,-1,1,-1),\nn.
\eea
The resulting mirror geometry is a copmlex three dimension hypersurface $P(W_3)$  embedded in $WP^4$ and can be shown to define a local Calabi-Yau threefold. The corresponding mirror potential $P(W_3)= W_3(x_1,x_2,x_3,x_4,w)$   reads as:
 \be
 W_3(x_1,x_2,x_3,x_4,w)=  x^4_1+ x^4_2+x^4_2+x^4_3+x^4_4+x_1x_2x_3x_4+ w^2( x^2_1+ x^2_2+x^2_2+x^2_3+x^2_4)+ w^2.
\ee
  $P(W_3) $ describes  a quasihomogenous hypersurface in $ WP^4_{1,1,1,1,2}$ with non zero first Chern class $ c_1$. We can restore the Calabi-Yau condition $ c_1=0$ by considering   $wP(W_3) $ as our hypersurface instead of eq(7.11), which describe a singular Calabi-Yau threefolds. Note in passing that $w$  dependant terms in eq(7.11) are exacly the monomials appearing in the standard mirror tetravalent geometry eq(7.8).
\section{ Conclusion}
 In this paper we have studied two main ideas: \\
 (1) We have completed partial results, obtained earlier in  physics literature, for embedding supersymmetric QFT with non simply laced gauge groups in superstring and F-theory compactification on local Calabi-Yau manifolds. In our study we have:\\
(a) shown that in general there exists two classes of solutions for the constraints eqs, of folding by outer-automorphisms of the ADE dynking diagrams. These classes  depend on   the various  realisations of the outer-automorphism goups of the Dynkin ADE when acting on the toric graph $\Delta$(ADE)of the mirror geometry of the blown up ADE singularities of ALE space.  \\
(b) given explicit results for affine non simply laced BCFG toric data and the corresponding mirror geometries for F-theory compactifications on elliptically fibered Calabi-Yau manifolds. These results extend known ones obtained in literature for the affine ADE cases.\\
(c) completed the analysis for ordianry ADE singularities by giving the explict derivation of the lacking non simply ones in type II superstring compactifications on local Calabi-Yau threefolds.\\
(2) We have also developed the basis of  polyvalent geometry. The latter extends naturally the known bivalent and trivalent geometries, used in the geometric engineering of supersymmetic (confomal) QFT embedded in superstring compactifications on Calabi-Yau manifolds with ordinary and affine singularities. This extented geometry is expected to be used for dealing with the analysis of singularities of higher dimensional toric varieties. In addition to this, polyvalent geometry find a remarkable application in the tetravalent case where it is used to derive a new  solution for the mirror geometry of affine $\hat D_4$ singularity. The new toric representation of the affine so(8) algebra is worked out explicitly using standard tetravalent geometry based on quartic lattice ${\bf Z}^4$ contrary to the special solution considerd in \cite{bfs1} based on a rational cubic lattice. The new solution for the mirror geometry of affine $D_4$ is a Calabi Yau hypersurface embedded in $WP^4(1,1,1,1,2)$.\\
\\
{\bf Acknowledgments}
This work is supported by SARS, programme de soutien a la recherche scientifique; Universit\'e Mohammed V-Agdal, Rabat.

\end{document}